\documentclass[runningheads]{llncs}

\usepackage[utf8]{inputenc}
\usepackage{color,soul}
\usepackage{xspace}
\usepackage{booktabs}
\usepackage{paralist}
\usepackage[group-separator={,}]{siunitx}
\usepackage{graphicx}
\usepackage{amsmath,amssymb}
\usepackage{multirow}
\usepackage{subfig}
\usepackage{epsfig}
\usepackage{float}
\usepackage{pgf}
\usepackage{tikz}
\usepackage{wrapfig}
\usepackage{url}
\usepackage{lipsum}

\usetikzlibrary{shapes.geometric,arrows,positioning,calc,fit,decorations.markings,backgrounds,shapes,shapes.multipart,shapes.geometric,backgrounds,automata}

\tikzset{->-/.style={thick,decoration={
  markings,
  mark=at position #1 with {\arrow{>}}},postaction={decorate}},
  rect/.style n args={4}{
        draw=none,
        rectangle,
        append after command={
            \pgfextra{%
                \pgfkeysgetvalue{/pgf/outer xsep}{\oxsep}
                \pgfkeysgetvalue{/pgf/outer ysep}{\oysep}
                \def\arg@one{#1}
                \def\arg@two{#2}
                \def\arg@three{#3}
                \def\arg@four{#4}
                egin{pgfinterruptpath}
                    \ifx\\#1\\\else
                        \draw[draw,#1] ([xshift=-\oxsep,yshift=+\pgflinewidth]\tikzlastnode.south east) edge ([xshift=-\oxsep,yshift=0\ifx\arg@two\@empty-\pgflinewidth\fi]\tikzlastnode.north east);
                    \fi\ifx\\#2\\\else
                        \draw[draw,#2] ([xshift=-\pgflinewidth,yshift=-\oysep]\tikzlastnode.north east) edge ([xshift=0\ifx\arg@three\@empty+\pgflinewidth\fi,yshift=-\oysep]\tikzlastnode.north west);
                    \fi\ifx\\#3\\\else
                        \draw[draw,#3] ([xshift=\oxsep,yshift=0-\pgflinewidth]\tikzlastnode.north west) edge ([xshift=\oxsep,yshift=0\ifx\arg@four\@empty+\pgflinewidth\fi]\tikzlastnode.south west);
                    \fi\ifx\\#4\\\else
                        \draw[draw,#4] ([xshift=0+\pgflinewidth,yshift=\oysep]\tikzlastnode.south west) edge ([xshift=0\ifx\arg@one\@empty-\pgflinewidth\fi,yshift=\oysep]\tikzlastnode.south east);
                    \fi
                \end{pgfinterruptpath}
            }
        }
  },
  rect'/.style n args={4}{
        rectangle,
        append after command={
            \pgfextra{%
                \pgfkeysgetvalue{/pgf/outer xsep}{\oxsep}
                \pgfkeysgetvalue{/pgf/outer ysep}{\oysep}
                \begin{pgfinterruptpath}
                    \ifx\\#1\\\else
                        \draw[draw,#1] ([xshift=-\oxsep,yshift=0]\tikzlastnode.south east) edge ([xshift=-\oxsep,yshift=0]\tikzlastnode.north east);
                    \fi\ifx\\#2\\\else
                        \draw[draw,#2] ([xshift=-\pgflinewidth,yshift=-\oysep]\tikzlastnode.north east) edge ([xshift=0+\pgflinewidth,yshift=-\oysep]\tikzlastnode.north west);
                    \fi\ifx\\#3\\\else
                        \draw[draw,#3] ([xshift=\oxsep,yshift=0-\pgflinewidth]\tikzlastnode.north west) edge ([xshift=\oxsep,yshift=0+\pgflinewidth]\tikzlastnode.south west);
                    \fi\ifx\\#4\\\else
                        \draw[draw,#4] ([xshift=0+\pgflinewidth,yshift=\oysep]\tikzlastnode.south west) edge ([xshift=0-\pgflinewidth,yshift=\oysep]\tikzlastnode.south east);
                    \fi
                \end{pgfinterruptpath}
            }
        }
  },
  dontshortenme/.style={
        shorten >=0pt,
        shorten <=0pt
  },
  rect''/.style n args={4}{
        draw=none,
        rectangle,
        append after command={
            \pgfextra{%
                \pgfkeysgetvalue{/pgf/outer xsep}{\oxsep}
                \pgfkeysgetvalue{/pgf/outer ysep}{\oysep}
                \def\my@path{\path[shorten >=\pgflinewidth,shorten <=\pgflinewidth] ([xshift=-\oxsep]\tikzlastnode.south east) edge}
                \def\arg@{#1}
                \ifx\arg@\@empty
                    \def\arg@{draw=none}
                \fi
                \eappto\my@path{[\arg@] }
                \appto\my@path{ ([xshift=-\oxsep]\tikzlastnode.north east)
                                          ([yshift=-\oysep]\tikzlastnode.north east) edge }
                \def\arg@{#2}
                \ifx\arg@\@empty
                    \def\arg@{draw=none}
                \fi
                \eappto\my@path{[\arg@] }
                \appto\my@path{ ([yshift=-\oysep]\tikzlastnode.north west)
                                          ([xshift=\oxsep] \tikzlastnode.north west) edge }
                \def\arg@{#3}
                \ifx\arg@\@empty
                    \def\arg@{draw=none}
                \fi
                \eappto\my@path{[\arg@] }
                \appto\my@path{ ([xshift=\oxsep]\tikzlastnode.south west)
                                          ([yshift=\oysep] \tikzlastnode.south west) edge }
                \def\arg@{#4}
                \ifx\arg@\@empty
                    \def\arg@{draw=none}
                \fi
                \eappto\my@path{[\arg@] }
                \appto\my@path{ ([yshift=\oysep]\tikzlastnode.south east);}
                \begin{pgfinterruptpath}
                    \my@path
                \end{pgfinterruptpath}
            }
        }
  }
}

\newcommand\blfootnote[1]{%
  \begingroup
  \renewcommand\thefootnote{}\footnote{#1}%
  \addtocounter{footnote}{-1}%
  \endgroup
}

\newcommand{\sbip}{\mathcal{S}\text{BIP}}

\begin{document}

\title{Performance Evaluation of the NDN Data Plane Using Statistical Model Checking}
\titlerunning{NDN Performance Evaluation using SMC}
%

\author{Siham Khoussi\inst{1,2} \and Ayoub Nouri\inst{1}\and Junxiao Shi\inst{2}  \\
James Filliben\inst{2} \and Lotfi Benmohamed\inst{2} \and Abdella Battou\inst{2} \and Saddek Bensalem\inst{1}}

\authorrunning{S. Khoussi et al.}

\institute{Univ. Grenoble Alpes, CNRS, Grenoble INP\footnote{Institute of Engineering Univ. Grenoble Alpes}, VERIMAG, 38000 Grenoble, France \and
National Institute of Standards and Technology, Gaithersburg, MD, 20899, USA.
}
\maketitle

\begin{abstract}
Named Data Networking (NDN) is an emerging internet architecture that addresses weaknesses of the Internet Protocol (IP). Since Internet users and applications have demonstrated an ever-increasing need for high speed packet forwarding, research groups have investigated different designs and implementations for fast NDN data plane forwarders and claimed they were capable of achieving high throughput rates. However, the correctness of these statements is not supported by any verification technique or formal proof. In this paper, we propose using a formal model-based approach to overcome this issue. We consider 
the NDN-DPDK prototype implementation of a forwarder developed at the National Institute of Standards and Technology (NIST), which leverages concurrency to enhance overall quality of service. 
We use our approach to improve its design and to formally demonstrate that it can achieve high throughput rates.
\end{abstract}

\thispagestyle{empty}

\section{Introduction}\label{sec:intro}
\blfootnote{The identification of any commercial product or trade name does not imply endorsement or recommendation by the National Institute of Standards and Technology, nor is it intended to imply that the materials or equipment identified are necessarily the best available for the purpose.}
With the ever growing number of communicating devices, their intensive information usage and the increasingly critical security issues, research groups have recognized the limitations of the current Internet architecture based on The internet protocol (IP) \cite{NDN-vision}. Information-Centric Networking (ICN) is a new paradigm 
that transforms the Internet from a host-centric paradigm, as we know it today, to an end-to-end paradigm focusing on the content, hence more appropriate to our modern communication practices. It promises better security, 
mobility and scalability. 

Several research projects grew out of ICN. Examples include content-centric architecture, Data Oriented Network Architecture and many others \cite{6563278}, but one project stood out the most and was sponsored by the National Science Foundation (NSF) called Named Data Networking (NDN) \cite{Zhang:2014:NDN:2656877.2656887}. NDN is gaining rapidly in popularity and has even started being advertised by major networking players \cite{footnote}.


IP was designed to answer a different challenge, that is of creating a communication network, where packets named only communication endpoints. The NDN project proposes to generalizes this setting, such that packets can name other objects, i.e. \emph{``NDN changes the semantics of network services from delivering the packet to a given destination address to fetching data identified by a given name. The name in an NDN packet can name anything \-- an endpoint, a data chunk in a movie or a book, a command to turn on some lights, etc.``}~\cite{Zhang:2014:NDN:2656877.2656887}. This simple change
has deep implications in term of routers forwarding performance since data needs to be fetched 
from an initially unknown location.


Being a new concept, NDN (Section~\ref{sec:ndn}) has not undergone any formal verification work yet. The initial phase of the project was meant to come up with proof-of-concept prototypes for the proposed architecture. This has lead to a plethora of less performing implementations in terms of packets' forwarding rates (throughput). A lot of effort was then directed to optimizing NDN forwarders' performances by trying different data structures (Hash maps) and targeting different hardware (GP-GPU). Unfortunately, validation was mainly carried using pure simulation and testing techniques.


In this work, we take a step back and try to tackle the performance problem differently. We consider a model-based approach that allows for rigorous reasoning and formal verification (Section~\ref{sec:approach}). In particular, we rely on the $\sbip$ framework \cite{DBLP:conf/atva/MediouniNBDLB18,doi:10.1504/IJCCBS.2018.096439} offering a stochastic component-based modeling formalism and Statistical Model Checking (SMC) engine. $\sbip$ is used along an iterative and systematic design process which consists of four phases (1) building a parameterized functional system model, which does not include performance (2) run a corresponding implementation in order to collect context information and performance measurements, characterized as probability distribution functions, (3) use these distributions to create a stochastic timed performance model and (4) use SMC to verify that the obtained model satisfies requirements of interest. 

This approach is applied to verify that the NDN Data Plane Development Kit (NDN-DPDK) (an effort to develop a high performance forwarder for NDN networks at the National Institute of Standards and Technology (NIST)) can perform at high packet forwarding rates (Section~\ref{sec:design}). We investigate different design alternatives regarding concurrency (number of threads), system dimensioning (queues sizes) and deployment (mapping threads to multi-core). Using our approach, we were able to figure out what are the best design parameters to achieve higher performances (Section~\ref{sec:analysis}). These were taken into account by the NDN developers at NIST to enhance the ongoing design and implementation. To the best of our knowledge, this is the first work using formal methods in the context of the NDN project.
\vspace{-.4cm}

\section{Named Data Networking}\label{sec:ndn}
\vspace{-.25cm}
This section describes the NDN protocol and introduces the NDN-DPDK forwarder being designed and implemented at NIST.
\subsection{Overview}
NDN is a new Internet architecture different from IP. Its core design is exclusively based on naming contents rather than end points (IP addresses in the case of IP) and its routing is based on name prefix lookups \cite{Jacobson2009}.

The protocol supports three types of packets, namely \emph{Interest, Data} and \emph{Nack}. Interests are consumer requests sent to a network and Data packets are content producers replies. The Nack lets the forwarder know of the network's inability to forward Interests further. One of NDN's advantages is its ability to cache content (Data) everywhere the Data packet propagates, making the NDN router stateful. Thus, future Interests are no longer required to fetch the content from the source, instead Data could be retrieved directly from a closer node that has a cached copy.

Packets in NDN travel throughout a network as follow: first a client application sends an Interest with a name prefix that represents the requested content. Names in NDN are hierarchical (e.g., /YouTube/Alex/video1.mpg denotes
 a YouTube video called Video1.mpg by a Youtuber Alex). Then, this packet is forwarded by the network nodes based on its name prefix. Finally, this Interest is satisfied with Data by the original source that produced this content or by intermediate routers that cached it due to previous requests. It is also crucial to note that consecutive transmissions of Interest packets with similar name prefix might not lead to the same path each time, but could rather be forwarded along different paths each time a request is made, depending on the forwarding strategy in place. This means that the same Data could originate from different sources (producers or caches). 

The NDN forwarding daemon (NFD) \cite{Afanasyev}, has three different data structures: \emph{Pending Interest Table (PIT), Content Store (CS)} and \emph{Forwarding Interest Base (FIB)}. The packet processing, according to the NDN protocol, is as follows (fig~\ref{fig:dataplane}):

\begin{figure}{}
\centering
\includegraphics[width=0.65\linewidth]{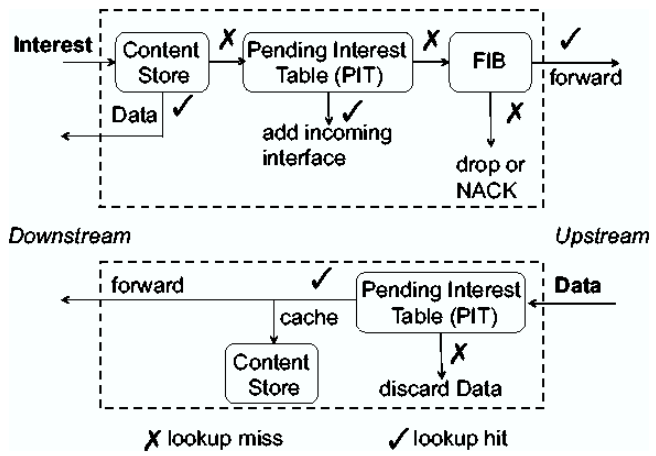}
\caption{NDN Data Plane}~\cite{plane} \label{fig:dataplane}
\vspace{-0.6cm}
\end{figure}

\begin{asparaenum}[1 --]
\item For Interests, the forwarder, upon receiving an Interest, starts off by querying the CS for possible copies of the Data, if a CS match is found during this operation, the cached Data is returned downstream towards the client. Otherwise, an entry is created in the PIT with its source and destination faces (communication channels that the forwarder uses for packet forwarding) for record keeping. Using the PIT, the forwarder determines whether the Interest is looped in the network by checking a global unique number called Nonce in the Interest against existing previous PIT entries. If a duplicate nonce is found the Interest is dropped and a Nack of reason \emph{Duplicate} is sent towards the requester. Otherwise, the FIB is queried for a possible next hop to forward the Interest towards an upstream node; if there is no FIB match, the Interest is immediately dropped and replied with a Nack of reason \emph{No Route}.
\item For Data, the forwarder starts off by querying the PIT. If a PIT entry is found, the Data is sent to downstream nodes listed in the PIT entry, then the PIT arms a timer to signal the deletion of this entry and a copy of the Data is immediately stored in the CS for future queries. 
If no record is found in the PIT, the Data is considered malicious and discarded. 
\end{asparaenum}

\subsection{The NDN-DPDK Forwarder}
NDN-DPDK is a forwarder developed at NIST to follow the NDN protocol and to leverage concurrency. In this paper, we evaluate its capacity to achieve high throughput using Statistical Model Checking (SMC).

\begin{figure}[ht]
\vspace{-0.4cm}
\centering
\includegraphics[width=0.98\linewidth]{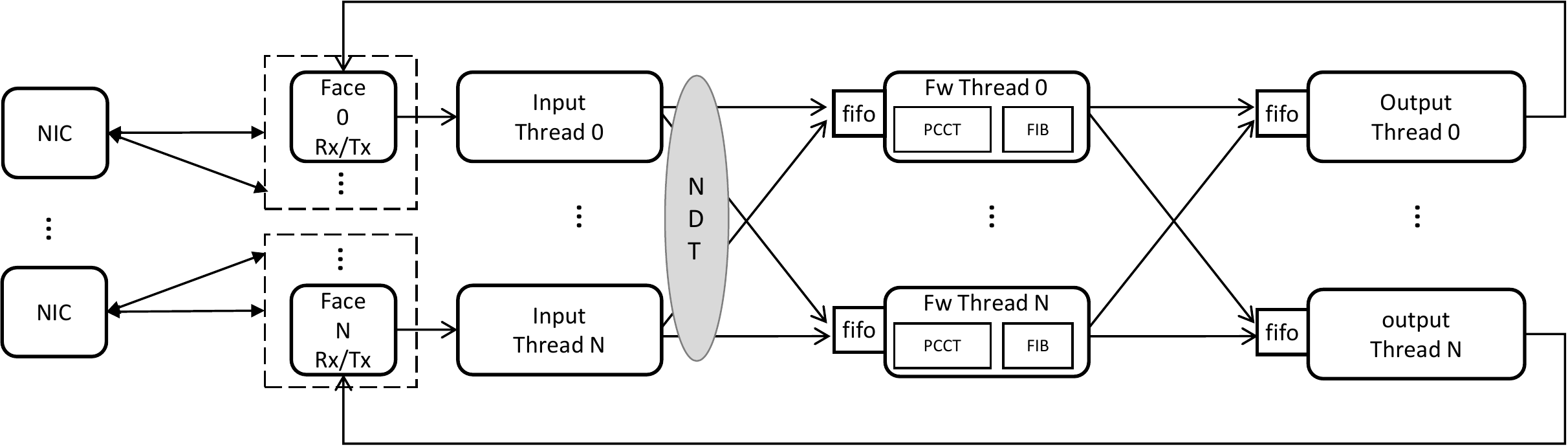}
\caption{Diagram of the NDN-DPDK forwarder}\label{fig:ndnfw-dpdk-diagram}
\end{figure}

The NDN-DPDK forwarder's data plane has three stages: input, forwarding, and output (Fig.~\ref{fig:ndnfw-dpdk-diagram}).
Each stage is implemented as one or more threads pinned to CPU cores, allocated during initialization.
\textbf{Input} threads receive packets from a Network Interface Card (NIC) through faces, decode them, and dispatch them to forwarding threads.
The \textbf{forwarding} thread processes Interest, Data, or Nack packets according to the NDN protocol. \textbf{Output} threads send packets via faces then queue them for transmission on their respective NIC.

\vspace{-0.4cm}
During forwarder initialization, each hardware NIC is provided with a large memory pool to place incoming packets.
The input thread continuously polls the NIC to obtain bursts of \num{64} received packets. Then decodes, reassembles fragmented packets, and drops malformed ones.
Then, it dispatches each packet to the responsible forwarding thread which is determined as follows:
\begin{inparaenum}[(a)]
\item For an Interest, the input thread computes  SipHash of its first two name components and queries the last 16 bits of the hash value in the Name Dispatch Table (NDT), a \num{65536} entry lookup table configured by the operator, to select the forwarding thread.
\item Data and Nack carry a 1-byte field in the packet header which indicates the forwarding thread that handled the corresponding Interest. Once identified, Data (or Nack) will be dispatched to the same one.
\end{inparaenum}

The forwarding thread receives packets dispatched by input threads through a queue.
It processes each packet according to the NDN protocol, using two data structures both implemented as hash tables:
\begin{inparaenum}[(a)]
\item The FIB records where the content might be available and which forwarding strategy is responsible for the name prefix.
\item The PIT-CS Composite Table (PCCT) records which downstream node requested a piece of content, and also serves as a content cache; it combines the PIT and CS found in a traditional NDN forwarder.
\end{inparaenum}
%
%

The output thread retrieves outgoing packets from forwarding threads through a queue. Packets are fragmented if necessary and queued for transmission on a NIC. The NIC driver automatically frees the memory used by packets after their transmission, making it available for newly arrived packets.

\section{Formal Model-based Approach}\label{sec:approach}
In this section, we describe the methodology used in this study which includes the underlying modeling formalism as well as the associated analysis technique.

\subsection{Overview}
\begin{figure}[ht]
\vspace{-0.4cm}
\centering
\includegraphics[width=0.9\linewidth]{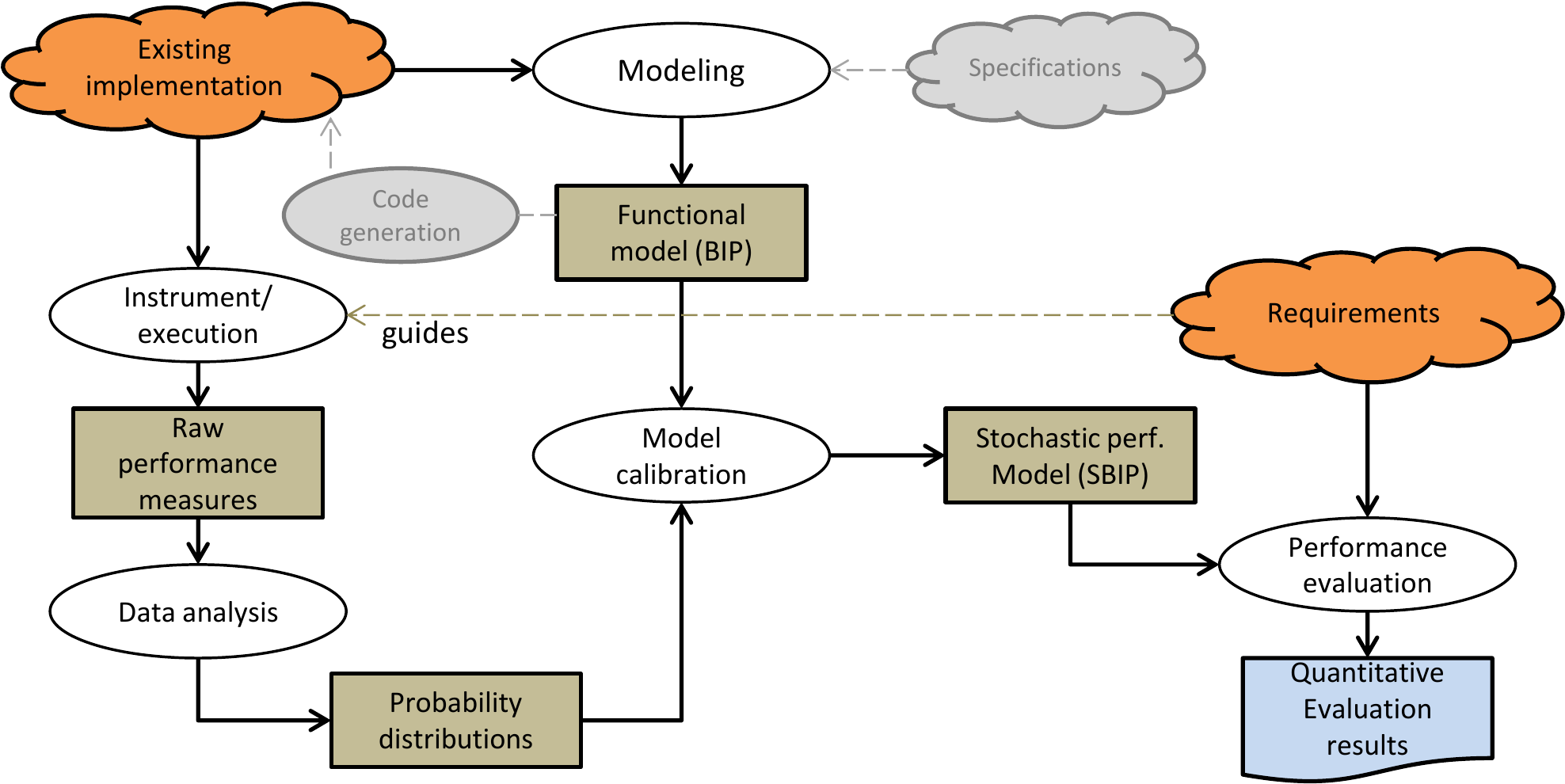}
\caption{Performance evaluation approach for NDN data plane.}\label{fig:approach}
\vspace{-0.5cm}
\end{figure}

Our methodology (Fig.~\ref{fig:approach}) is based on a formal model. In order to evaluate a system's performance, its model must be faithful, i.e. it must reflect the real characteristics and behavior of this system. Moreover, to allow for exhaustive analyses, this model needs to be formally defined and the technique used for analysis needs to be trustworthy and scalable. Our approach adheres to these principles in two ways. First, by relying on the $\sbip$ formal framework (introduced below) that encompasses a stochastic component-based modeling formalism and an SMC engine for analysis  \cite{DBLP:conf/atva/MediouniNBDLB18}. Second, by providing a method for systematically building formal stochastic models for verification that combine accurate performance information with the functional behavior of the system.

\begin{figure}[ht]
\centering
\includegraphics[width=0.65\linewidth]{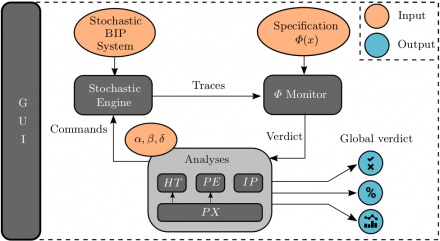}
\caption{SBIP engine \cite{nouri_2017}}\label{fig:sbipo}
\end{figure}

This approach takes a functional system model and a set of requirements to verify. The functional model could be obtained from a high-level specification or an existing implementation (we use the latter in this paper). The system's implementation which could also be obtained by automatic code generation, is instrumented and used to collect performance measurements regarding the requirements of interest, e.g. throughput. These measurements are analyzed and characterized in the form of probability density functions with the help of statistical techniques such as sensitivity analysis and distribution fitting. The obtained probability density functions are then introduced in the functional model using a well defined calibration procedure \cite{DBLP:journals/tecs/NouriBMLB16}. The latter produces a stochastic timed model (when measurements concern time), which will be analyzed using the SMC engine depicted in fig~\ref{fig:sbipo} \cite{nouri_2017}.
Note that the considered models in this approach or workflow can be parameterized with respect to different aspects that we want to analyze and explore. Basically, the defined components types are designed to be instantiated in different context, e.g. with different probability density functions thus showing different performance behaviors.
While, the model considered for analysis using SMC is a specific instance for which all the parameters are fixed, some degree of parameterization is still allowed on the verified requirements. 

\subsection{Stochastic Component-based Modeling in BIP}
\label{sec:sbip-model}
BIP (Behavior, Interaction, Priority) is a highly expressive component based framework for rigorous system design \cite{Basu-Bozga-Sifakis-06}. It allows the
construction of complex, hierarchically structured models from atomic components 
characterized by their behavior and their interfaces. Such components are
transition systems enriched with variables. Transitions are used to move from
a source to a destination location. Each time a transition is taken, component
variables may be assigned new values, computed by user-defined C/C++ functions. Composition of BIP components is expressed by layered application of interactions and priorities. Interactions express synchronization constraints between actions of the composed components while priorities are used to filter among possible interactions e.g. to express scheduling policies.

The stochastic semantics of BIP were initially introduced in \cite{Nouri:2015:SMC:2744465.2744509} and recently extended for real-time systems in \cite{doi:10.1504/IJCCBS.2018.096439}. They enable the definition of 
\begin{wrapfigure}[12]{r}{.47\textwidth}
\vspace{-0.3cm}
 \centering
 \scalebox{0.8}{
     \begin{tikzpicture}[->,>=stealth',shorten >=1pt,auto,semithick]
 		\node[state] (s0) {$s_0$};
 		\coordinate[left=5ex of s0] (i);
 		\node[state, right=6.4ex and 15ex of s0] (s1) {$s_1$};
			
 		\path (i) edge node[above, xshift=-.3cm] {$t=0$} (s0)
 			  (s0) edge[bend left] node[above] {$p\vartriangleright$} (s1)
 			       edge[loop above] node[above] {$\mathbf{recv}$} ()
 			  (s1) edge[bend left] node[below, align=left, xshift=-.4cm] {$[t=p]$\\$\mathbf{snd}$} (s0)
 				   edge[loop below] node[left, align=right, xshift=-.1cm, yshift=-.3cm] {$[t<p]$\\$\mathbf{tick}$\\$t++$} ()
 				   edge[loop above] node[above] {$\mathbf{recv}$} ();

 	\end{tikzpicture}}
 	\vspace{-0.3cm}
 	\caption{A stochastic BIP component; client behavior issuing requests each time unit $p$.}
 	\label{fig:dtmc}
 \end{wrapfigure}
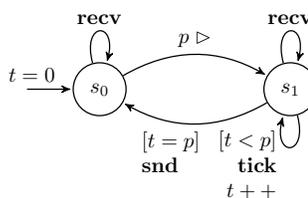
 stochastic components encompassing probabilistic variables updated according to user-defined probability distributions. The underlying mathematical model behind this is a Discrete Time Markov Chain. These are modeled as classical BIP components augmented with probabilistic variables as shown in Fig.~\ref{fig:dtmc} and depicts a client behavior in a client-server setting where the client issues a request (\textbf{snd}) each \textit{p} time units. The period \textit{p} is set probabilistically by sampling a distribution function ($p\vartriangleright$) given as a parameter of the model. Time is introduced by explicit \textbf{tick} transitions and waiting is modeled by exclusive guards on the \textbf{tick} and \textbf{snd} transitions with respect to time (captured in this example by the variable $t$).

\subsection{Statistical Model Checking in a Nutshell}
Statistical model-checking ({\em SMC}) \cite{HLMP04,You05a} is a formal verification method that combines simulation with statistical reasoning to provide quantitative answers on whether a stochastic system satisfies some requirements. It was successfully used in various domains such as biology~\cite{David2015}, communication~\cite{BBBCDL10}
and avionics~\cite{afdx}. 
It has the advantage to be applicable to models and implementations (provided that they meet specific assumptions) in addition to capturing rare events. The $\sbip$ SMC engine \cite{DBLP:conf/atva/MediouniNBDLB18} implements well-known statistical algorithms for stochastic systems verification, namely, Hypothesis Testing \cite{You05a}, Probability Estimation \cite{HLMP04} and Rare Events. In addition, it provides an automated parameters exploration procedure. 
The tools take as inputs a stochastic BIP model, a Linear-time/Metric Temporal Logic (LTL/MTL) property to check and a set of confidence parameters required by the statistical test.

\section{NDN-DPDK Modeling}\label{sec:design}
In this section we present the modeling process of the NDN-DPDK from a functional to a stochastic timed model for throughput evaluation.
\subsection{A Parameterized Functional BIP Model}
\begin{figure}[ht]
\centering
\includegraphics[width=0.95\linewidth]{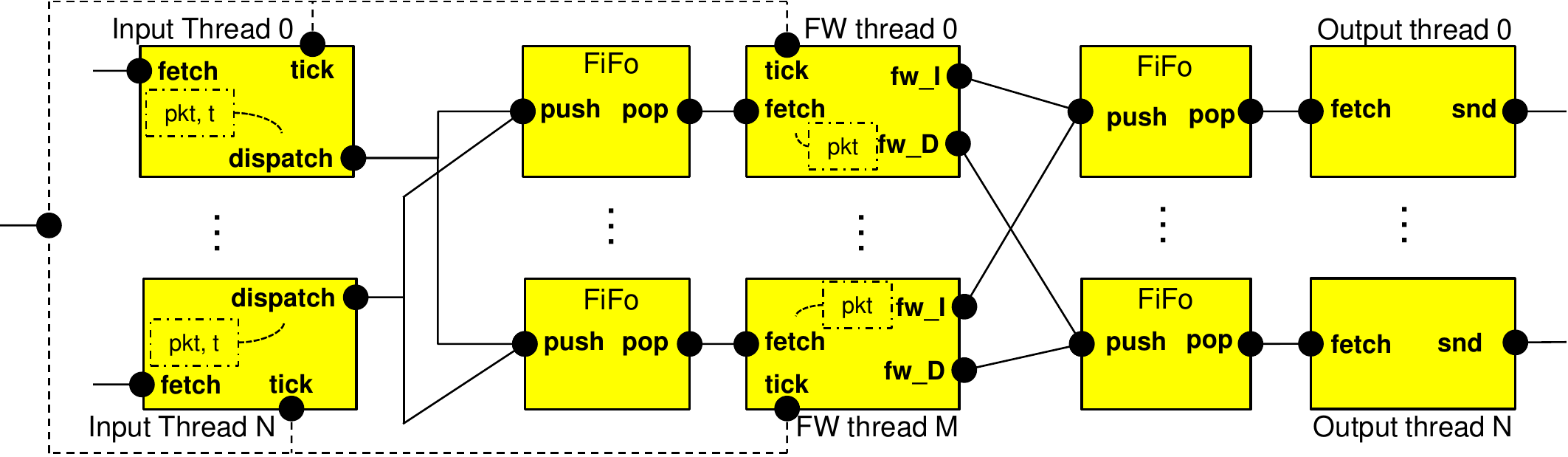}
\caption{A functional BIP model of the NDN-DPDK forwarder}\label{fig:bipm}

\end{figure}

Fig.~\ref{fig:bipm} depicts the BIP model of the NDN-DPDK forwarder introduced in Section~\ref{sec:ndn} which shows its architecture in terms of interacting BIP components that can easily be matched to the ones in Fig.~\ref{fig:ndnfw-dpdk-diagram}. The presented model is parameterized with respect to the number of components, their mapping into specific CPU cores, FIFOs sizes, etc. Due to space limitation, we present in~\cite{techRep} the behaviors of all the components of the NDN-DPDK forwarder in Fig.~\ref{fig:bipm}.
It is worth mentioning that the model is initially purely functional and untimed. Time is introduced later through the calibration procedure.

\subsection{Building the Performance Model}
\begin{figure}
    \centering
    \subfloat[Bloc diagram\label{fig:bloc-topo}]{
		\scalebox{0.7}{
		    \includegraphics[width=.5\textwidth]{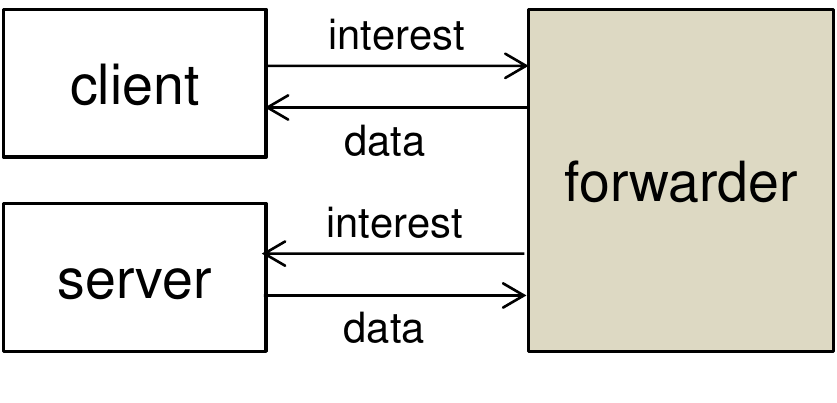}
		}
	} 
	\quad
	\subfloat[BIP model\label{fig:topology}]{
		\scalebox{0.9}{
		    \includegraphics[width=.5\textwidth]{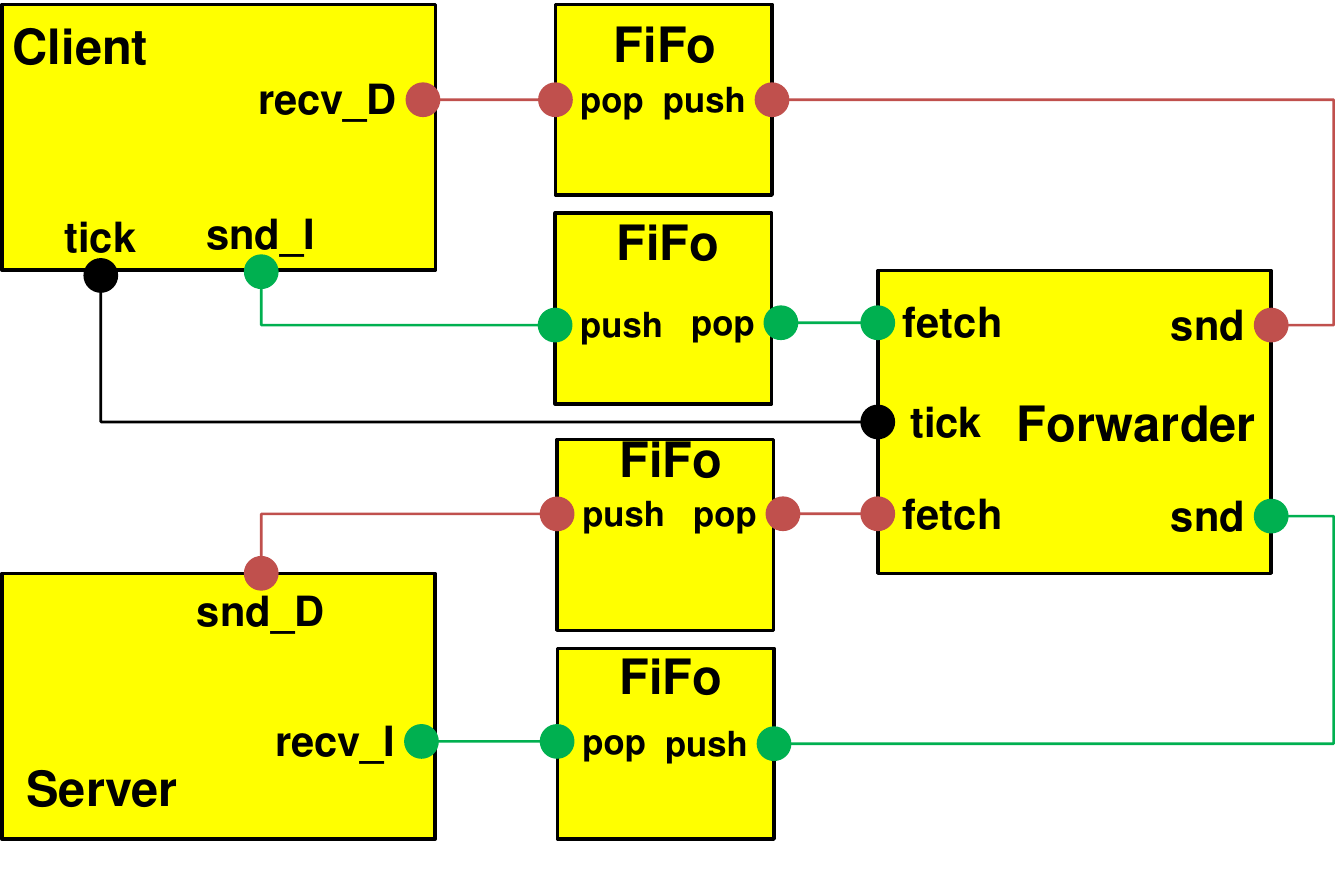}
		}
	} \label{fig:bip-topo}
	\caption{Considered network topology for the use case}
\end{figure}

To build the performance model that will be the subject of this analysis, we consider the network topology shown in Figure~\ref{fig:bloc-topo}.

The latter has a traffic generator client (consumer), a forwarder (NDN-DPDK), and a traffic generator server (producer), arranged linearly. Fig~\ref{fig:topology} corresponds to the BIP model of the topology where the green line indicates Interest packets' path from the client to the producer through the forwarder and the red line indicates Data packets' path towards the client. 
Fig~\ref{fig:consumer} shows the behavior of the consumer component in the BIP model of the topology (fig~\ref{fig:topology}) which is to send Interests at each time interval $\delta$ and receive Data then count the number of satisfied Interests in packets per second (pps). Whereas fig~\ref{fig:producer} shows the behavior of the producer component which is to simply generate and send a Data packet each time an Interest arrives.
\begin{figure}
\centering
\vspace{-0.7cm}
\begin{minipage}{.5\textwidth}
  \centering
   \begin{tikzpicture}[->,>=stealth',shorten >=1pt,auto,semithick]
 		\node[state] (s0) {$s_0$};
 		\coordinate[above=5ex of s0] (i);
 		\node[state, right=6.4ex and 15ex of s0] (s1) {$s_1$};
 		\node[state, left=6.4ex and 15ex of s0] (s2) {$s_2$};
 		
 		\path (i) edge node[left,align=center,pos=0, xshift=.5cm, yshift=0.3cm] {$t=0$} (s0)
 			  (s0) edge[bend left] node[below] {$\mathbf{}$} (s1)
 			       edge[bend left] node[below] {$\mathbf{recv\_D}$} (s2)
 			  (s1) edge[bend left] node[below, align=left, xshift=-.1cm] {$[t=\delta]$\\$\mathbf{send\_I}$} (s0)
 				   edge[loop below] node[left, align=right, xshift=.5cm, yshift=-.7cm] {$[t<\delta]$\\$\mathbf{tick}$\\$t++$} ()
 			  (s2) edge[bend left] node[above, align=left, xshift=-.1cm] {$Sat\_I++$} (s0);
 	\end{tikzpicture}
  \captionof{figure}{Consumer Model}
  \label{fig:consumer}
\end{minipage}%
\begin{minipage}{.5\textwidth}
  \centering
  \begin{tikzpicture}[->,>=stealth',shorten >=1pt,auto,semithick]
	\node[state] (s0) {$s_0$};
  		\coordinate[left=5ex of s0] (i);
  		\node[state, below=5ex of s0] (s1) {$s_1$};
  		\node[state, below=5ex of s1] (s2) {$s_2$};

	    \path (i) edge node[xshift=-.3cm] {$t=0$} (s0)
	         (s0) edge[loop right] node[left, align=right, xshift=.6cm, yshift=.9cm] {$\mathbf{tick}$\\$t++$} ()
	              edge[] node[right] {$\mathbf{recv\_I}$} (s1)
			 (s1) edge[] node[right] {$\mathbf{gen\_D}$} (s2)
			 (s2) edge[bend left] node[above, xshift=-.7cm] {$\mathbf{send\_D}$} (s0);
 \end{tikzpicture}
  \captionof{figure}{Producer model}
  \label{fig:producer}
\end{minipage}
\end{figure}

The structure of the NDN-DPDK model (Figure~\ref{fig:bipm}) calls for four distribution functions to characterize performance:
\begin{inparaenum}[a)]
\item Interest dispatching latency in input threads.
\item Data dispatching latency in input threads.
\item Interest forwarding latency in forwarding threads.
\item Data forwarding latency in forwarding threads.
\end{inparaenum}
Notice that Nack packets are out of the scope of these experiments. We identified the following factors that can \textit{potentially} affect the system's performance:
\begin{compactenum}[1.]
\item \label{stochastic-factor-nfwds}
      \textbf{Number of forwarding threads.}
      Having more forwarding threads distributes workload onto more CPU cores. The cores can compete for the shared L3 cache, and potentially increase forwarding latency of individual packets.
\item \label{stochastic-factor-numa}
      \textbf{Placement of forwarding threads onto Non Uniform Memory Access nodes (NUMA).}
      Input threads and their memory pools are always placed on the same NUMA node as the Ethernet adapter whereas the output threads and the forwarding threads can be moved across the two nodes.
      If a packet is dispatched to a forwarding thread on a different node, the forwarding latency is generally higher because memory access is crossing NUMA boundaries. 
\item \label{stochastic-factor-ncomps}
      \textbf{Packet name length measured by the number of its components.}
      A longer name requires more iterations during table lookups, potentially increasing Interest forwarding latency.
\item \label{stochastic-factor-payloadlen}
      \textbf{Data payload length.}
      Although the Data payloads are never copied, a higher payload length increases demand for memory bandwidth, thus potentially increasing latencies.
\item \label{stochastic-factor-pps}
      \textbf{Interest sending rate from the client.}
      Higher sending rate requires more memory bandwidth, thus potentially increasing latencies.
      It may also lead to packet loss if queues between input and forwarding threads overflow.
\item \label{stochastic-factor-pit}
      \textbf{Number of PIT entries.}
      Although the forwarder's PIT is a hash table that normally offers $O(1)$ lookup complexity, a large number of PIT entries inevitably leads to hash collisions, which could increase forwarding latency.
\item \textbf{Forwarding thread's queue capacity.}
      the queues are suspected to impact the overall throughput of the router through packet overflow and loss rates. However, it does not influence packets individual latencies.\label{fact7}
\end{compactenum}

After identifying the factors with potential influence on packet latency, we instrument the real forwarder to collect latency measurements. Then, perform statistical analysis to identify which factors are more significant. This narrows down the number of factors used and associated distribution functions. 
\vspace{-.2cm}
\subsubsection{Forwarder Instrumentation.}\label{sec:stochastic-instrument} 
Factors~\ref{stochastic-factor-nfwds},~\ref{stochastic-factor-numa},~\ref{stochastic-factor-ncomps},~\ref{stochastic-factor-payloadlen}, ~\ref{stochastic-factor-pps} and \ref{fact7} can be controlled by adjusting the forwarder and traffic generator configuration, while factor~\ref{stochastic-factor-pit} is a result of network traffic and is not in our control.
To collect the measurement, we modified the forwarder to log packets latencies as well as the PIT size after each burst of packets.
We minimized the extra work that input threads and forwarding threads have to perform to enable instrumentation, leaving the measurement collection to a separate logging thread or post-processing scripts. It is important to mention that this task does in fact introduce timing overhead. Therefore, the values obtained will have a bias (overestimate) that translates into additional latency but the trends observed remain valid. 

We conducted the experiment on a Supermicro server equipped with two Intel E5-2680V2 processors, \num{512} GB DDR4 memory in two channels, and four Mellanox ConnectX-5 100 Gbit/s Ethernet adapters.
The hardware resources are evenly divided into two NUMA nodes.
To create the topology in Fig.~\ref{fig:topology}, we connected two QSFP28 passive copper cables to connect the four Ethernet adapters and form two point-to-point links.
All forwarders and traffic generator processes were allocated with separate hardware resources and could only communicate over Ethernet adapters.

In each experiment, the consumer transmitted either at sending intervals of one Interest per \num{700} ns or per \num{500} ns under 255 different name prefixes.
There were 255 FIB entries registered in the NDN-DPDK forwarder at runtime (one for each name prefix used by the consumer), all of which pointed to the producer node.
The producer would reply to every Interest with a Data packet of the same name.
The forwarder's logging thread was configured to discard the first $67\,108\,864$ samples (either latency trace or PIT size) during warm-up period, and then collect the next $16\,777\,216$ samples and ignore the cool down session.
Each experiment represents about 4 million Interest-Data exchanges.
\begin{table}[h]
\centering
\caption{Factors used. NUMA mapping is described below.}\label{table:stochastic-factor-range}
\begin{tabular}{|c|c|c|c|c|}
\hline
Factors & forwarding threads & Name length & Payload length             & Sending intervals \\ \hline
Values  & \{1, 2, 3, 4, 5, 6, 7, 8\}   & \{3, 7, 13\}             & \{0, 300, 600, 900, 1200\} & \{500 ns, 700 ns\}         \\ \hline
\end{tabular}
\vspace{-0.5cm}
\end{table}
We repeated the experiment using different combinations of the factors in Table~\ref{table:stochastic-factor-range} and the following NUMA arrangements:
\begin{itemize}[]
\item{(P1)} Client and server faces and forwarding threads are all on the same NUMA,\label{best}
\item{(P2)} Client face and forwarding threads on one NUMA, server face on the other,\label{bx}
\item{(P3)} Client face on one NUMA, forwarding threads and server face on the other,\label{c}
\item{(P4)} Client face and server face on one NUMA, forwarding threads on the other.\label{worst}
\end{itemize}

In P1, packet latency is expected to be the smallest because all processes are placed on the same NUMA therefore, no inter-socket communication and no overhead are introduced. In P4, both Interests and Data packets are crossing NUMA boundaries twice since the forwarding threads are pinned to one NUMA whereas the client and the server faces, connected to the Ethernet adapters, reside on another. This is suspected to increase packet latency tremendously as opposed to P1, P2 and P3.
These suspicions predict that placement P1 is the best case scenario and placement P4 is obviously the worst. However, we aim at getting more insight and confidence through quantitative formal analysis. This will provide a recommendation as to which placement is better suited based on the remaining parameters combinations.

\subsubsection{Model Fitting.}\label{sec:fitting}
\begin{figure}[ht]
\centering
\includegraphics[width=0.8\linewidth]{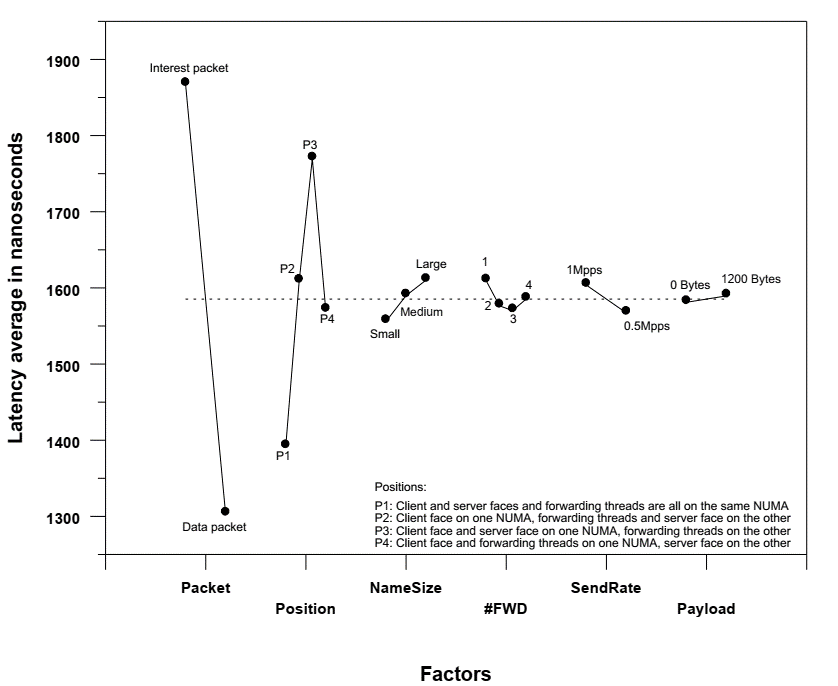}
\caption{Main Effects Plot for Interest and Data packets}
\label{fig:dex}
\vspace{-0.4cm}
\end{figure}

Before calibrating our functional BIP model with multiple distinctive probability distributions representing each combination of the factors, we choose to reduce the number of used distributions by performing a sensitivity analysis. This analysis examines the impact of several factors on the response (packet latency) and discovers the ones that are more important. In this paper, we use DataPlot~\cite{dataplot} to produce the Main Effect Plot (Fig.~\ref{fig:dex}) for factors
\ref{stochastic-factor-nfwds} to 
\ref{stochastic-factor-pps}.

The plot shows steeper line slopes for the packet type (packet type is not a factor. We intend to show how the NDN-DPDK forwarder processes both Interest and Data differently) as well as factors (\ref{stochastic-factor-nfwds}), 
(\ref{stochastic-factor-numa}), (\ref{stochastic-factor-ncomps}), and (\ref{stochastic-factor-pps}) which indicates a greater magnitude of the main effect on the latency. However, it shows almost a horizontal line for factor~\ref{stochastic-factor-payloadlen} inducing an insignificant impact on the latency. The latter is explained by the fact that the forwarder processes packet names (headers) only and doesn't read Data payloads.
As for the PIT size (factor~\ref{stochastic-factor-pit}), it is expected to heavily increase packet latency when it is full. However, because this table's implementation is optimized for high performance and entries are continuously removed when Data packets arrive (PIT entries being satisfied), we confirmed through a correlation analysis that we can ignore this factor's impact.

Based on the analysis above, we build distribution functions for each of the factors that have greater impacts on packet latency in this study. These factors are: 
\begin{inparaenum}[]
\item{(1)} the number of forwarding threads,
\item{(2)} NUMA placement,
\item{(3)} packet name size (header),
\item{(5)} sending rate and,
\item{(7)} FIFO capacity (FIFO impacts the loss rates and not individual packet latency.
\end{inparaenum}
We refer the reader to \cite{techRep} to understand how we obtained the probability distributions for these factors.

\subsubsection{Probability distribution fitting.}\label{sec:fitting}
In order to determine a best fit distribution to a data set we follow the general procedure below using DataPlot~\cite{dataplot}:

\begin{enumerate}
    \item Determine if a normal distribution fits the data via a histogram or a normal probability plot or a normal probability plot correlation coefficient (ppcc);
    
    \item If the normal probability plot is not linear "enough" (that is, the normal probability plot correlation coefficient is not close enough to 1), we check the data against popular distributional families (e.g. Weibull ppcc plot, Gamma ppcc plot, Lognormal ppcc plot, etc);
    
    \item If none of the distributional families provide a ppcc value which is close "enough" to 1, then we apply a box-cox transformation to determine the best power that will make the transformed data normal, then we note the best power (lambda);
    
    \item Confirm the normality of the transformed data by a histogram, a normal probability plot or by a normal probability plot correlation coefficient (ppcc) \cite{Filliben};
    
    \item Identify the best location and scale parameter estimates for the transformed data using the transformed data's mean $\bar{y}$ and standard deviation $\bar{s}$;
    
    
    \item Generate the normal probability N(0,1) of a random sample (Z) of size n;
    
    \item Transform the N(0,1) random sample to a N($\bar{y}$,$\bar{s}$) random sample of the transformed data;
    
    \item Transform $ZT$ the random sample to the actual data random sample \cite{box-cox};
    
    \item Finally, reaffirm the similarity between $Y$ and $\bar{Y}$ by using a Bihistogram for example.

\end{enumerate}

\subsubsection{Model Calibration.}\label{sec:calibration}
Calibration is a well defined model transformation that transforms functional components into stochastic timed ones \cite{DBLP:phd/hal/Nouri15}.
In this section, we use the probability distributions obtained to calibrate the functional BIP model of the NDN-DPDK forwarder shown in Fig.~\ref{fig:bipm}. 

Figs.~\ref{fig:sbip-exp},~\ref{fig:input} and~\ref{fig:output} respectively show the  calibrated behaviors of a forwarding thread, an input thread customized to send traffic towards two forwarding threads (fwd=1, fwd=2) and an output thread. The forwarding time of Interest and Data packets is modeled using two probability distributions, $f_I$ and $f_D$ respectively. Whereas, the input threads' processing time of packets is modeled with the probability distribution $f_p$ which refers to either Interests or Data depending on where the component is located.

In the next section, we perform SMC on the calibrated model of the NDN-DPDK forwarder and explain the results.
\begin{figure}
\centering
\scalebox{0.9}{
		\begin{tikzpicture}[->,>=stealth',shorten >=1pt,auto,semithick]
			\node[state] (s0) {$f_0$};
			\coordinate[above left=5ex of s0] (i);
			\node[state, below=4ex and 15ex of s0] (s1) {$f_1$};
			\node[state, below left=6ex and 10ex of s1] (s2) {$f_2$};
			\node[state, below right=6ex and 10ex of s1] (s3) {$f_3$};
			
			\path (i) edge node[left,align=center,pos=0] {$t=0$} (s0)
						(s0) edge[] node[right] {$\mathbf{fetch}$} (s1)
						     edge[loop above] node[right, xshift=.1cm] {$\mathbf{tick}$} ()
						(s1) edge[bend right] node[below, align=center, xshift=.35cm, yshift=-.1cm] {$[pkt=D]$\\$f_D\vartriangleright$\\$fwd_D()$} (s2)
					 	     edge[bend left] node[below, align=center, xshift=-.35cm, yshift=-.1cm] {$[pkt=I]$\\$f_I\vartriangleright$\\$fwd_I()$} (s3)
					 	(s2) edge[bend left] node[left,align=center, xshift=-.1cm] {$[t=f_D]$\\$\mathbf{fw\_D}$} (s0)
						     edge[loop below] node[right,align=center, xshift=.1cm, yshift=-.2cm] {$[t<f_D]$\\$\mathbf{tick}$\\$t++$} ()
						(s3) edge[bend right] node[right,align=center, xshift=.1cm] {$[t=f_I]$\\$\mathbf{fw\_I}$} (s0)
						     edge[loop below] node[left,align=center, xshift=-.1cm, yshift=-.2cm] {$[t<f_I]$\\$\mathbf{tick}$\\$t++$} ();
						     
		\end{tikzpicture}} 
	\caption{Forwarding thread; forwarding time of Interest and Data is modeled using two distributions, resp. $f_I$, $f_D$.}
	\label{fig:sbip-exp}
\end{figure}
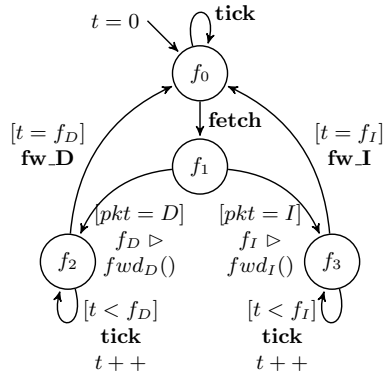
 
 \begin{figure}
\centering
\begin{minipage}{.5\textwidth}
  \centering
   			\begin{tikzpicture}[->,>=stealth',shorten >=1pt,auto,semithick]
			\node[state] (s0) {$f_0$};
			\coordinate[above left=5ex of s0] (i);
			\node[state, below=4ex and 15ex of s0] (s1) {$f_1$};
			\node[state, below left=6ex and 10ex of s1] (s2) {$f_2$};
			\node[state, below right=6ex and 10ex of s1] (s3) {$f_3$};
			
			\path (i) edge node[left,align=center,pos=0] {$t=0$} (s0)
						(s0) edge[] node[right] {$\mathbf{recv}$} (s1)
						     edge[loop above] node[right, align=right, xshift=-.5cm, yshift=.5cm] {$\mathbf{tick }$\\$t++$} ()
						(s1) edge[bend right] node[below, align=center, xshift=.4cm, yshift=-.35cm] {$[fwd=1]$\\$fp\vartriangleright$\\$in()$} (s2)
					 	     edge[bend left] node[below, align=center, xshift=-.4cm, yshift=-.35cm] {$[fwd=2]$\\$fp\vartriangleright$\\$in()$} (s3)
					 	     
					 	(s2) edge[bend left] node[left,align=center, xshift=-.2cm] {$[t=fp]$\\$\mathbf{to\_fwd1}$} (s0)
						     edge[loop below] node[right,align=center, xshift=-.7cm, yshift=-.7cm] {$[t<fp]$\\$\mathbf{tick}$\\$t++$} ()
						(s3) edge[bend right] node[right,align=center, xshift=.2cm] {$[t=fp]$\\$\mathbf{to\_fwd2}$} (s0)
						     edge[loop below] node[left,align=center, xshift=.7cm, yshift=-.7cm] {$[t<fp]$\\$\mathbf{tick}$\\$t++$} ();
						     
		\end{tikzpicture}
  \captionof{figure}{Input thread: processing time of packets is modeled using the distribution $f_p$ (p refers to Interests or data }
  \label{fig:input}
\end{minipage}%
\begin{minipage}{.5\textwidth}
  \centering
       \begin{tikzpicture}[->,>=stealth',shorten >=1pt,auto,semithick]
 		\node[state] (s0) {$s_0$};
 		\coordinate[left=5ex of s0] (i);
 		\node[state, right=6.4ex and 15ex of s0] (s1) {$s_1$};
 		\path (i) edge node[above, xshift=-.3cm] {$t=0$} (s0)
 			  (s0) edge[bend left] node[above] {$\mathbf{recv\_pkt}$} (s1)
 			       edge[loop above] node[above, align=right, xshift=.1cm, yshift=.1cm] {$\mathbf{tick}$\\$t++$} ()
 			  (s1) edge[bend left] node[below, align=left, xshift=-.1cm] {$\mathbf{send\_pkt}$} (s0)
 				   edge[loop below] node[left, align=right, xshift=.5cm, yshift=-.5cm] {$\mathbf{tick}$\\$t++$} ();
 	\end{tikzpicture}
  \captionof{figure}{Output thread mode}
  \label{fig:output}
\end{minipage}
\end{figure}

\section{Performance Analysis using SMC}\label{sec:analysis}
\subsection{Experimental Settings}
%
%
We run the SMC tests using the probability estimation algorithm (PE) with precision parameters $\alpha=0.1$ and $\delta=0.1$. Each test is configured with a different combination of values for the factors previously presented. And each execution of a test with a single set of parameters generates a single trace.
The property evaluated with the SMC engine is: \emph{Estimate the probability that all the issued Interests are satisfied, i.e. a Data is obtained in return for each Interest}. Thus, if PE concludes that the probability is 100 \%, then the verdict is positive.
\subsection{Analyses Results}

\subsubsection{Queues Dimensioning.}
First, we explore the impact of sizing forwarding threads queues. Each forwarding thread has an input queue. Initially, we consider a model with a single forwarding thread and vary its queue capacity with  \num{128}, \num{1024} or \num{4096} (in packets). Then set the client's sending rate to: $10^5$ packets per second (pps), $10^6$ pps or $10^7$ pps. The results are shown in Fig.~\ref{fig:queuesize}. The Y-axis represents the Interest satisfaction rate such that 100 \% (resp. 0 \%) indicates no loss (resp. 100 \% loss) and the x axis represents the queue capacity under different sending rates.
\begin{figure}[ht]
	\vspace{-0.5cm}
    \centering
    \subfloat[One Forwarding thread with different sending rates.]{
		    \includegraphics[width=.46\textwidth]{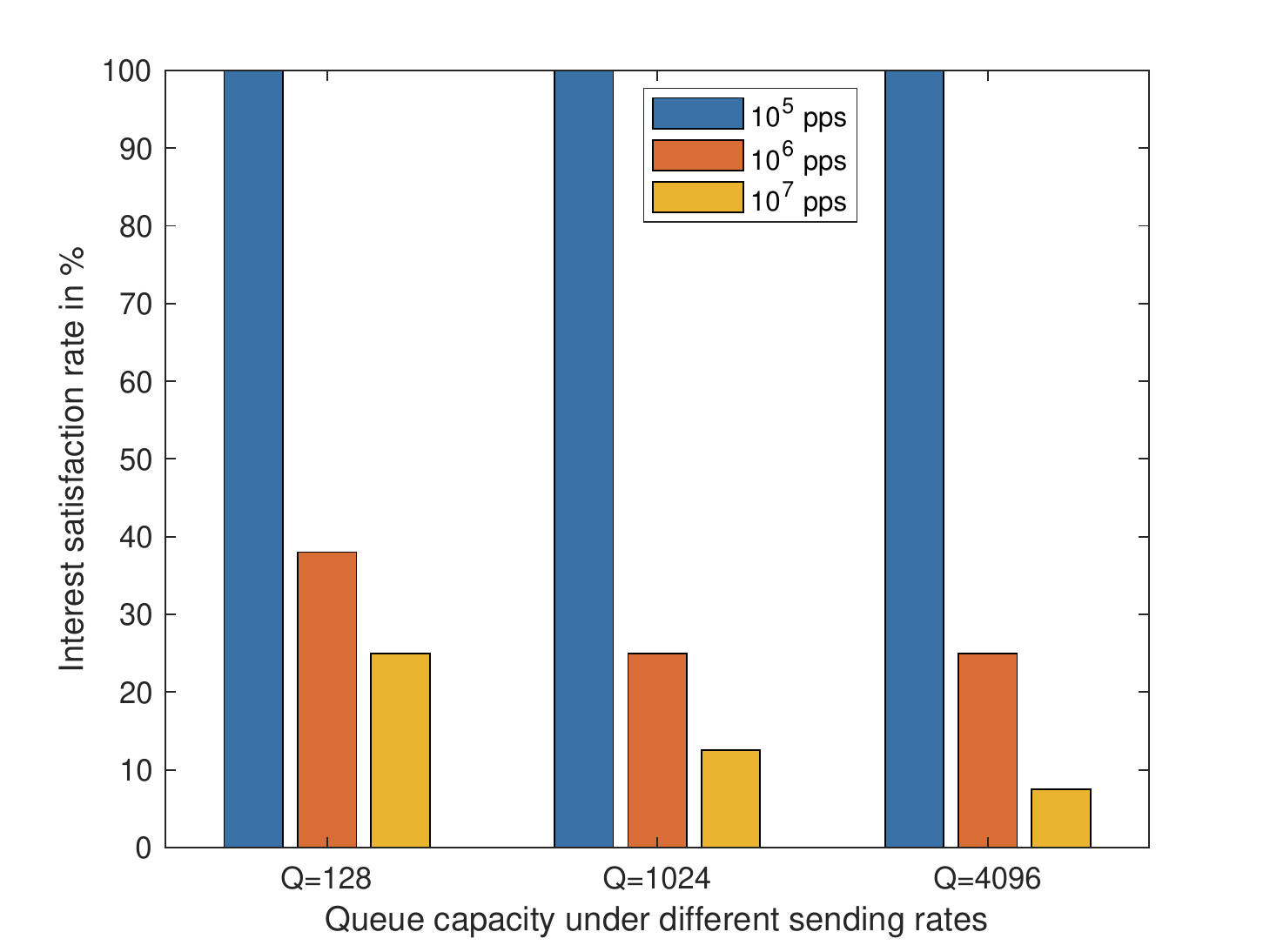}
		\label{fig:queuesize}
 	} 
	\quad
	\subfloat[Many Forwarding threads with a sending rate set to $10^6$ pps.]{
		    \includegraphics[width=.46\textwidth]{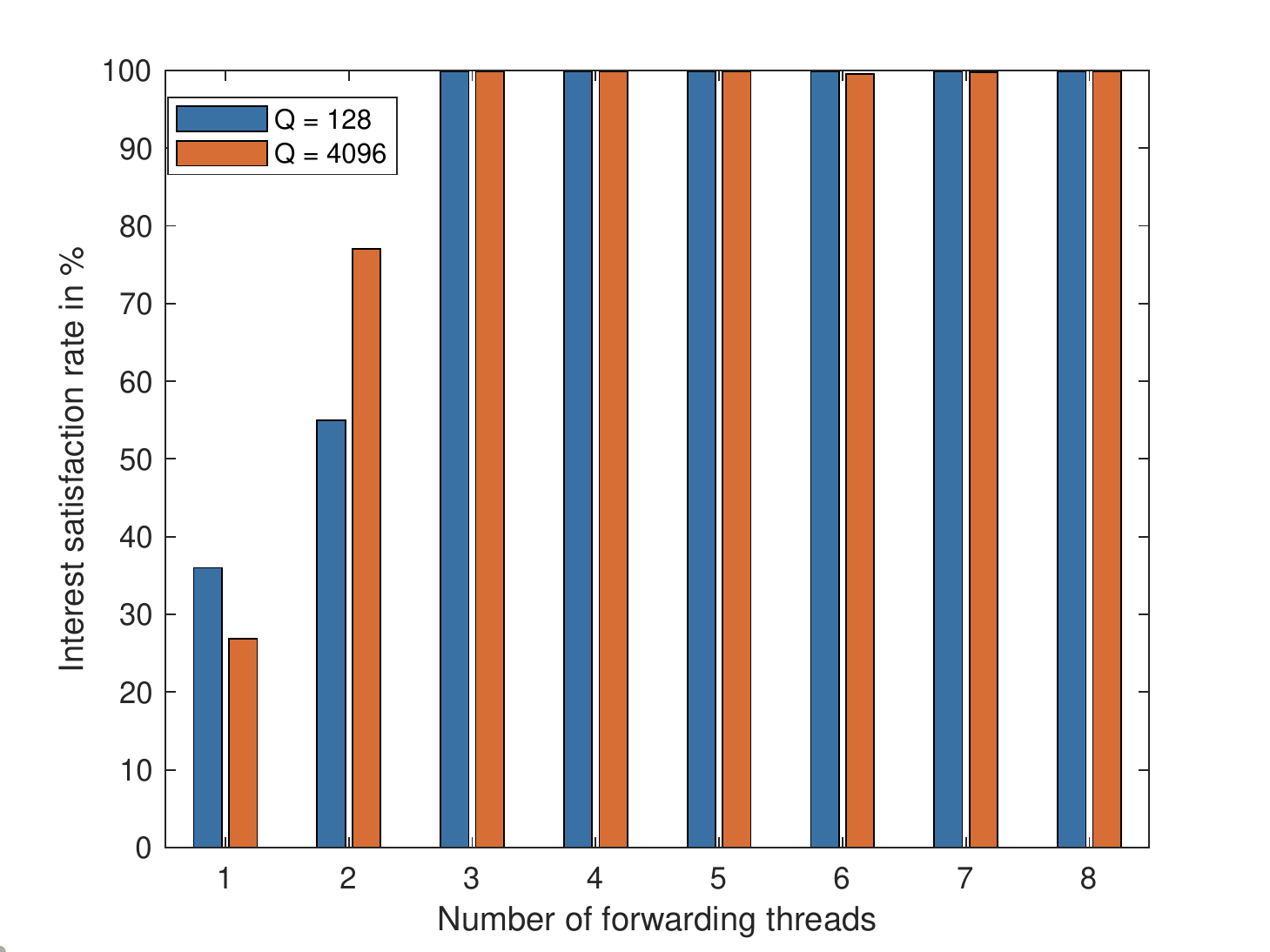}
		\label{fig:queueall}
	} 
	\caption{Exploration results of the Forwarding threads queues sizes.}
	\vspace{-0.6cm}
	\label{fig:queues}
\end{figure}

Fig.~\ref{fig:queuesize} indicates that at $10^5$ pps (blue), the Interest satisfaction rate is 100 \%. This means that the forwarder (with one forwarding thread) is capable of handling all packets at this sending rate ($10^5$ pps of packet size $1500$ bytes is equivalent to $1.2$ Gbps), under any queue size. However, under a faster sender rate (where a single forwarder shows signs of packet loss) we unexpectedly observed a better Interest satisfaction rate with a smaller queue (Q=$128$).
After a thorough investigation of the real implementation, we found out that the queues don't have proper management in terms of insertion and eviction policies that would give priority to Data over Interest packets. In the absence of such policy, more Interests would be queued while Data packets would be dropped resulting in Interests not being satisfied, thus lower performance (Interest satisfaction rate).
\emph{It is thus advised for the final implementation of the NDN-DPDK forwarder,  to use a queue capacity smaller than $128$ packets when the forwarder has a single forwarding thread and packets are sent at a fast rate.}

Similarly, we explore whether this observation remains true with more forwarding threads. In order to do that, we run SMC again on eight different models each with a different number of forwarding threads ($1$ to $8$) under a sending rate of $10^6$ pps ($1$ Interest per $1$ us) where a loss rate was observed in Fig.~\ref{fig:queuesize}. Then, we experimented with two queue capacities, namely $128$ and $4096$ packets. The results are reported in Fig.~\ref{fig:queueall}. The x Axis represents the number of forwarding threads while the y axis depicts the Interest satisfaction rate. 

 We observe that the queue size matters mainly in the case of a model with one and two forwarding threads. In fact, for a two threads model, a bigger queue size is preferred to maximize the performance, unlike when a single thread is used. As for the other six models, both sizes achieve almost 100 \% Interest satisfaction.
 This is due to the fact that three forwarding threads or more are capable of splitting the workload at $10^6$ pps and can pull enough packets from each queue with a minimum loss rate of 0.02 \% . \emph{This result stresses that, to avoid being concerned about a proper queue size, more threads are needed for handling a faster sending rate with minimum Interest loss.}

\subsubsection{NUMA placement, number of forwarding threads and packet name length.}
Another aspect to explore, is the impact of mapping the forwarding threads and/or NDN Faces to the two NUMA nodes (0, 1) under different sending rates and for multiple name lengths where Face 0 exchanges packets with the client and Face 1 with the server. To do that, we consider the four NUMA arrangements (P1), (P2), (P3) and (P4) in section \ref{sec:design} as well as the factors in Table~\ref{table:stochastic-factor-range} in the SMC analysis. 

In Figs.~\ref{s3} to \ref{f13}, each row represents experiments with similar packet name lengths \{small=3, medium=7, large=13\} and a queue capacity of 4096. The right-hand column indicates results for a faster sending rate of $2*10^6$ pps (500 ns interval) while the left-hand one shows results for a slower sending rate of $1.42*10^6$ pps (700 ns interval). 
The six figures includes four curves where each corresponds to the four NUMA arrangement options: P1 to P4.
\begin{figure}[htbp]
  \begin{minipage}[b]{0.52\linewidth}
    \centering
    \includegraphics[width=\linewidth]{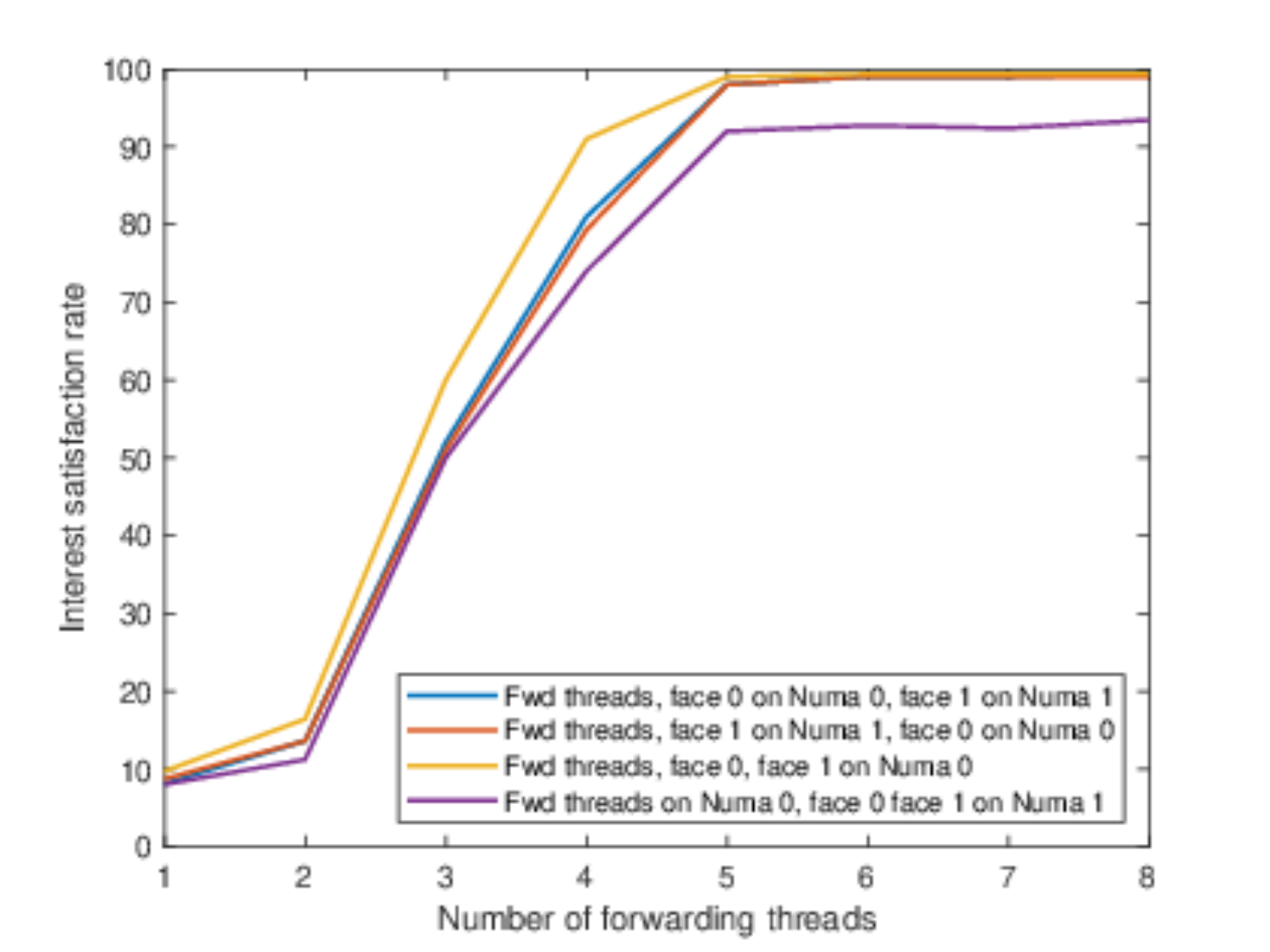}
    \caption{small names, 700 ns}\label{s3}
  \end{minipage}
  \begin{minipage}[b]{0.52\linewidth}
    \centering
    \includegraphics[width=\linewidth]{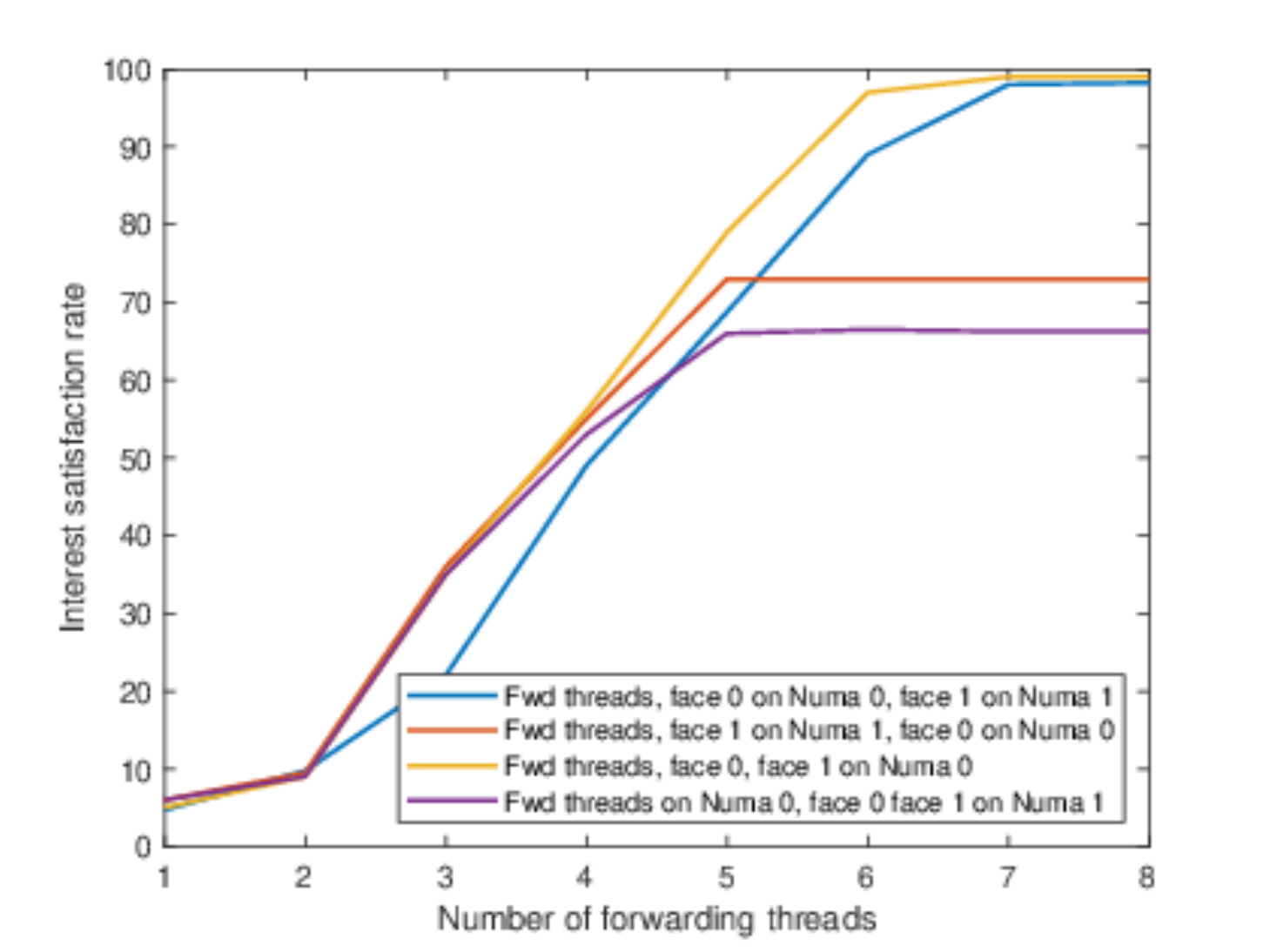}
    \caption{small names, 500 ns}\label{f3}
  \end{minipage}
    \begin{minipage}[b]{0.52\linewidth}
    \centering
    \includegraphics[width=\linewidth]{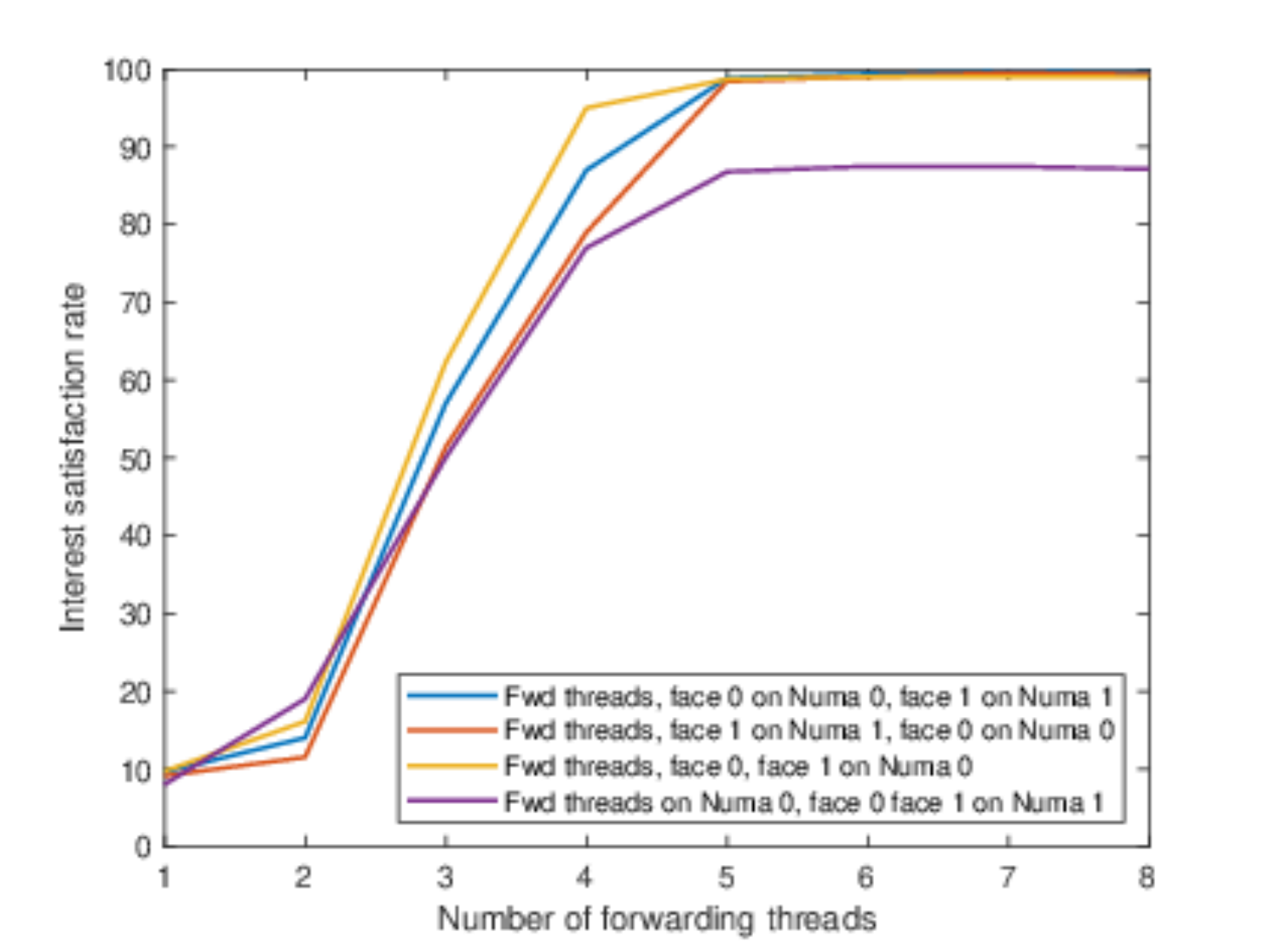}
    \caption{medium names, 700 ns}
    \label{s7}
  \end{minipage}
  \begin{minipage}[b]{0.52\linewidth}
    \centering
    \includegraphics[width=\linewidth]{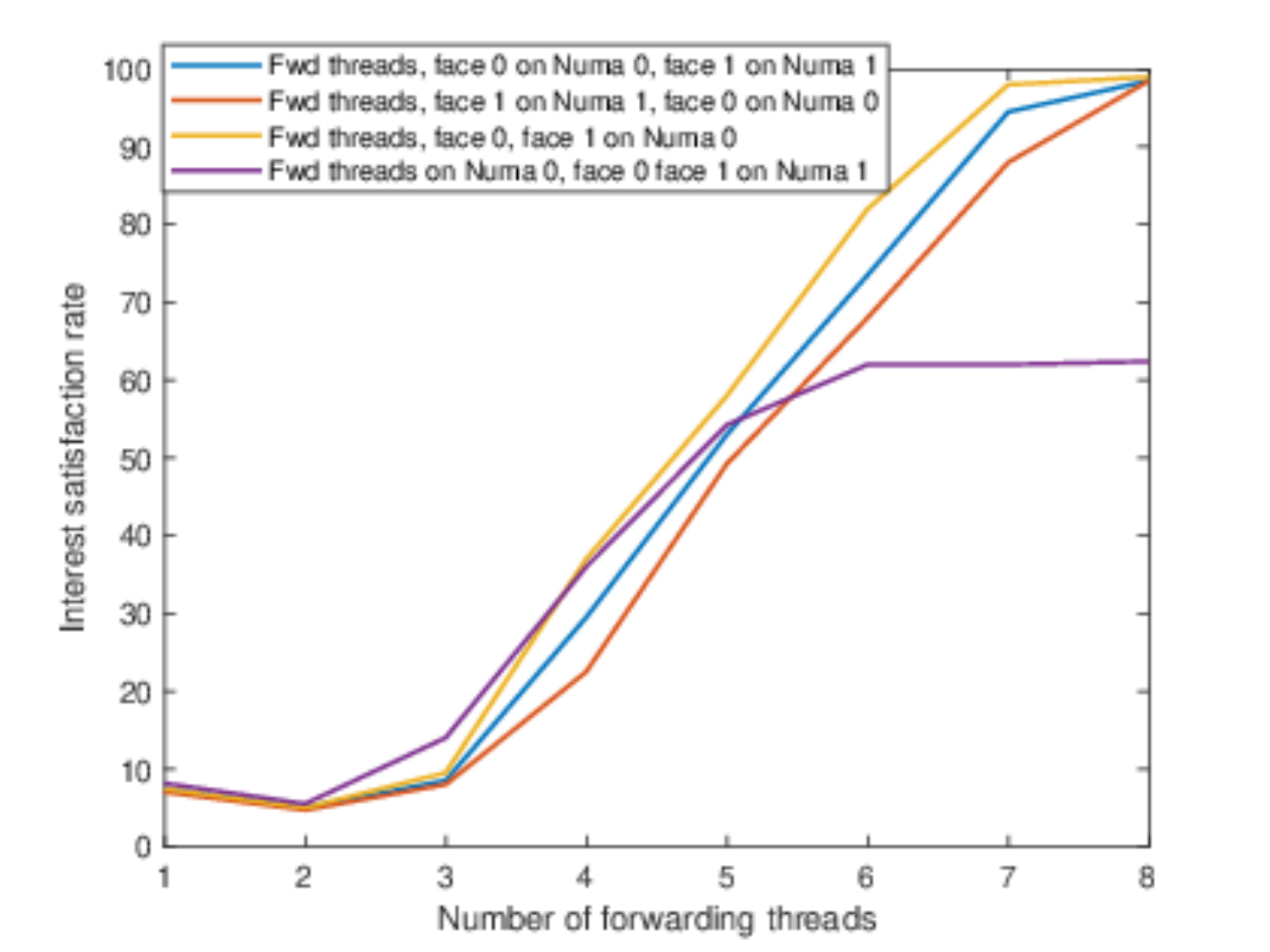}
    \caption{medium names, 500 ns}
    \label{f7}
  \end{minipage}
    \begin{minipage}[b]{0.52\linewidth}
    \centering
    \includegraphics[width=\linewidth]{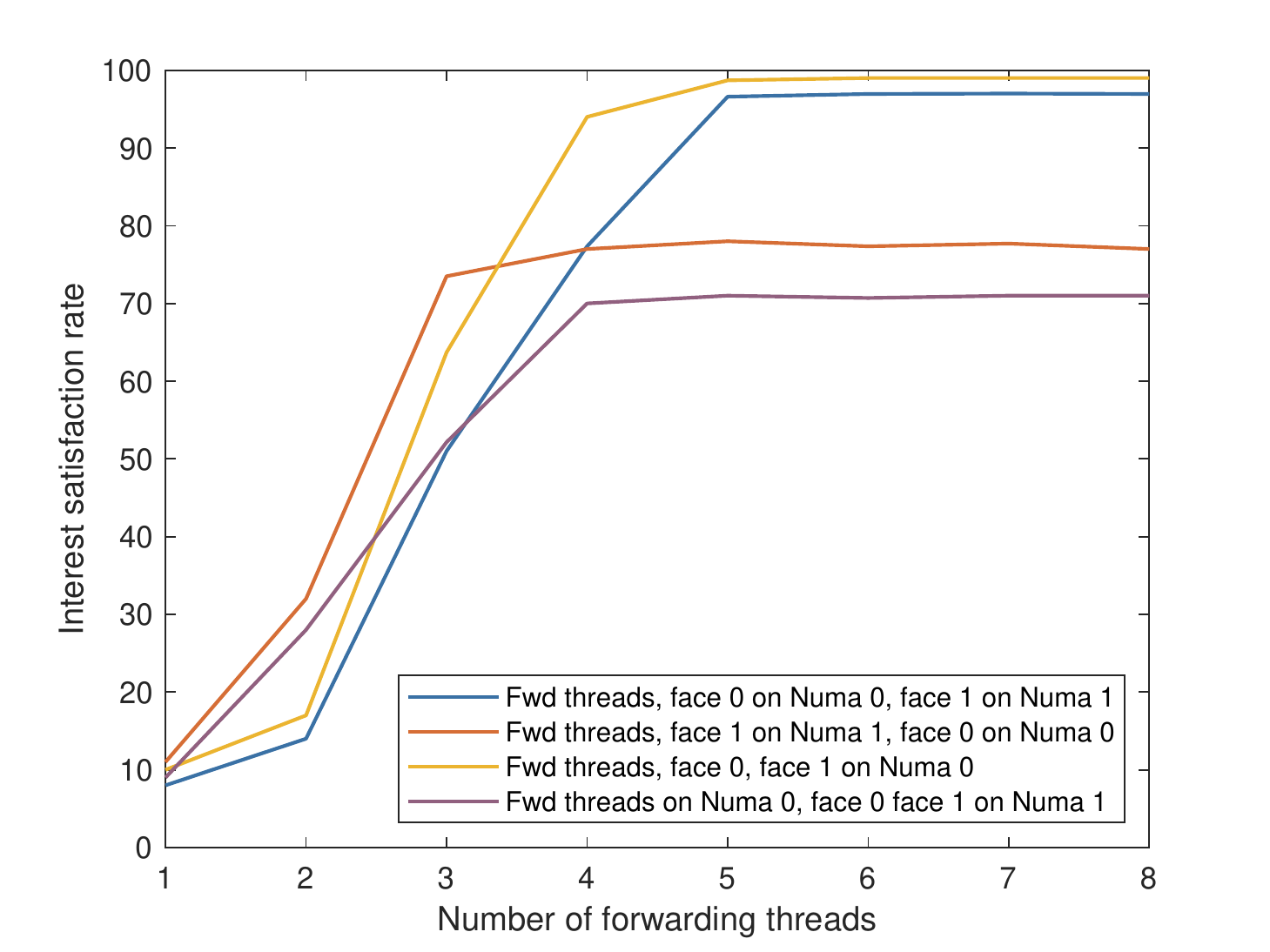}
    \caption{large names, 700 ns}
    \label{s13}
  \end{minipage}
  \begin{minipage}[b]{0.52\linewidth}
    \centering
    \includegraphics[width=\linewidth]{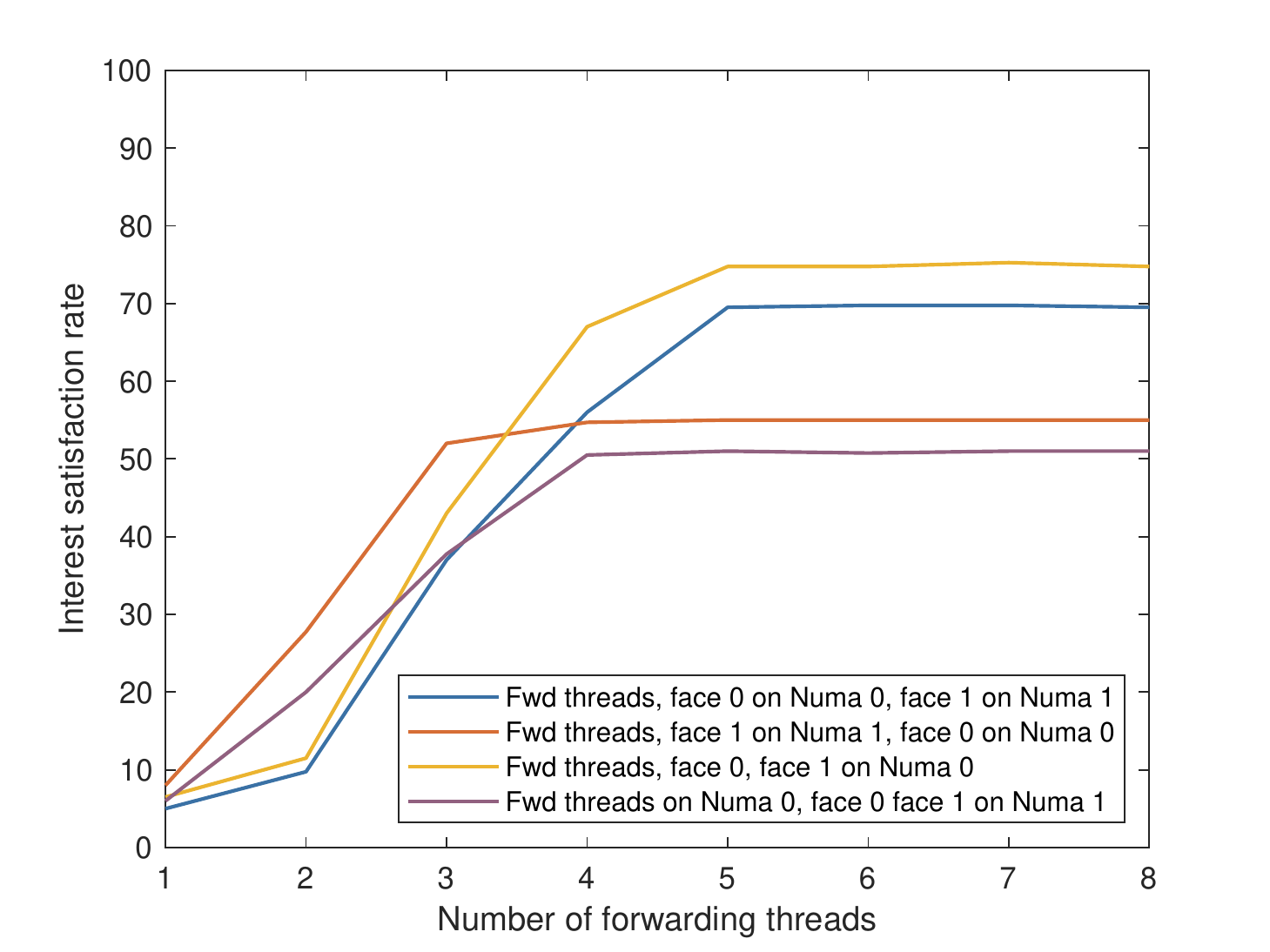}
    \caption{large names, 500 ns}
    \label{f13}
  \end{minipage}
\end{figure}

The six Figs.~\ref{s3} to \ref{f13} show that Interest satisfaction rates scale up with the increase of forwarding threads then reach a saturation plateau where adding more threads can no longer improve the performances.
Furthermore, with fewer forwarding threads, the loss rate is unavoidable and exceeds 80 \%. This is because the sending rate is faster than the forwarding threads processing capabilities causing their FIFO queues to saturate and start dropping packets frequently.
However, under a slower sending rate and packets with small, medium and large name lengths (3, 7, 13), Figs.~\ref{s3},~\ref{s7} and~\ref{s13} show that a maximum satisfaction rate of over 90 \% is achievable with only five forwarding threads.
Whereas when the client is generating packets faster at 2 Mpps, a saturation plateau of over $90$ \% is reached at six threads or more for small and medium names (Figs.~\ref{f3} and~\ref{f7}) and a plateau of slightly over $70$ \%, with five threads, for larger names (Fig.~\ref{f13}).
Also, Figs.~\ref{s3} and~\ref{s7} demonstrate that placing all processes (threads and faces) on a single NUMA (placement P1) outperforms the other three options. This observation is explained by the absence of inter-socket communication thus less timing overhead added such as in the case of the purple plot where packets are crossing NUMA boundaries twice from Face 0 to the forwarding threads then through Face 1 and back (placement P4).

\vspace{-0.03cm}
Figs.~\ref{f3} and~\ref{f7} show the impact of increasing the sending rate on packets with smaller names. In this case, it is preferred to also position all the processes on one NUMA such as the case of the yellow plot of the P1 series because NUMA boundary crossing usually downgrades the performance. In fact, the difference between no NUMA crossing and the double crossing (yellow and purple series respectively) is approximately 30 \% loss rate with more than five threads. The second best option P2 which is placing the forwarding threads on the NUMA receiving Interest packets with Face 0 (NUMA hosting the Ethernet adapter that receives Interests from the Client). However, when the number of threads is not in the saturation zone and the threads get overworked and start to loose packets, it is recommended to opt for placement P3. \emph{Based on these results, we recommend that for small to medium names, to use a maximum of eight threads but no less than five arranged as in placement P1 for optimum performances under a slower or a faster sending rate}. 

With a larger name however, Fig.~\ref{s13} depicts an unexpected behaviour when using three threads or less. In this case, placing the forwarding threads on the same NUMA as Face 1 (which is the Ethernet adapter connected to the server and receives Data packets), surpasses the other three options. Our explanation is that since forwarding threads take longer times to process incoming packets due to their longer name and timely lookup, particularly for Interests as they are searched by names inside the two tables (PCCT and FIB) rather than a token such as the case for Data packets. Placing the forwarding threads with the Data receiving Ethernet adapter connected to Face 1, has the potential to yield better results by quickly processing packets after a quick token search especially when the workload is bigger than the threads' processing capacity. When the sending rate is increased, the same results are observed in Fig.~\ref{f13} for a similar name length but with a decrease in performance. \emph{Thus, we recommend for larger names to use NUMA arrangement P3 only when the number of forwarding threads is less than three regardless of the sending rate (not advised due to high loss rate).}

\section{Lessons learned and future work}
\subsection{Experimental Settings}
%
%
We run the SMC tests using the probability estimation algorithm (PE) with precision parameters $\alpha=0.1$ and $\delta=0.1$. Each test is configured with a different combination of values for the factors previously presented. And each execution of a test with a single set of parameters generates a single trace.
The property evaluated with the SMC engine is: \emph{Estimate the probability that all the issued Interests are satisfied, i.e. a Data is obtained in return for each Interest}. Thus, if PE concludes that the probability is 100 \%, then the verdict is positive.
\subsection{Analyses Results}

\subsubsection{Queues Dimensioning.}
First, we explore the impact of sizing forwarding threads queues. Each forwarding thread has an input queue. Initially, we consider a model with a single forwarding thread and vary its queue capacity with  \num{128}, \num{1024} or \num{4096} (in packets). Then set the client's sending rate to: $10^5$ packets per second (pps), $10^6$ pps or $10^7$ pps. The results are shown in Fig.~\ref{fig:queuesize}. The Y-axis represents the Interest satisfaction rate such that 100 \% (resp. 0 \%) indicates no loss (resp. 100 \% loss) and the x axis represents the queue capacity under different sending rates.
\begin{figure}[ht]
	\vspace{-0.5cm}
    \centering
    \subfloat[One Forwarding thread with different sending rates.]{
		    \includegraphics[width=.46\textwidth]{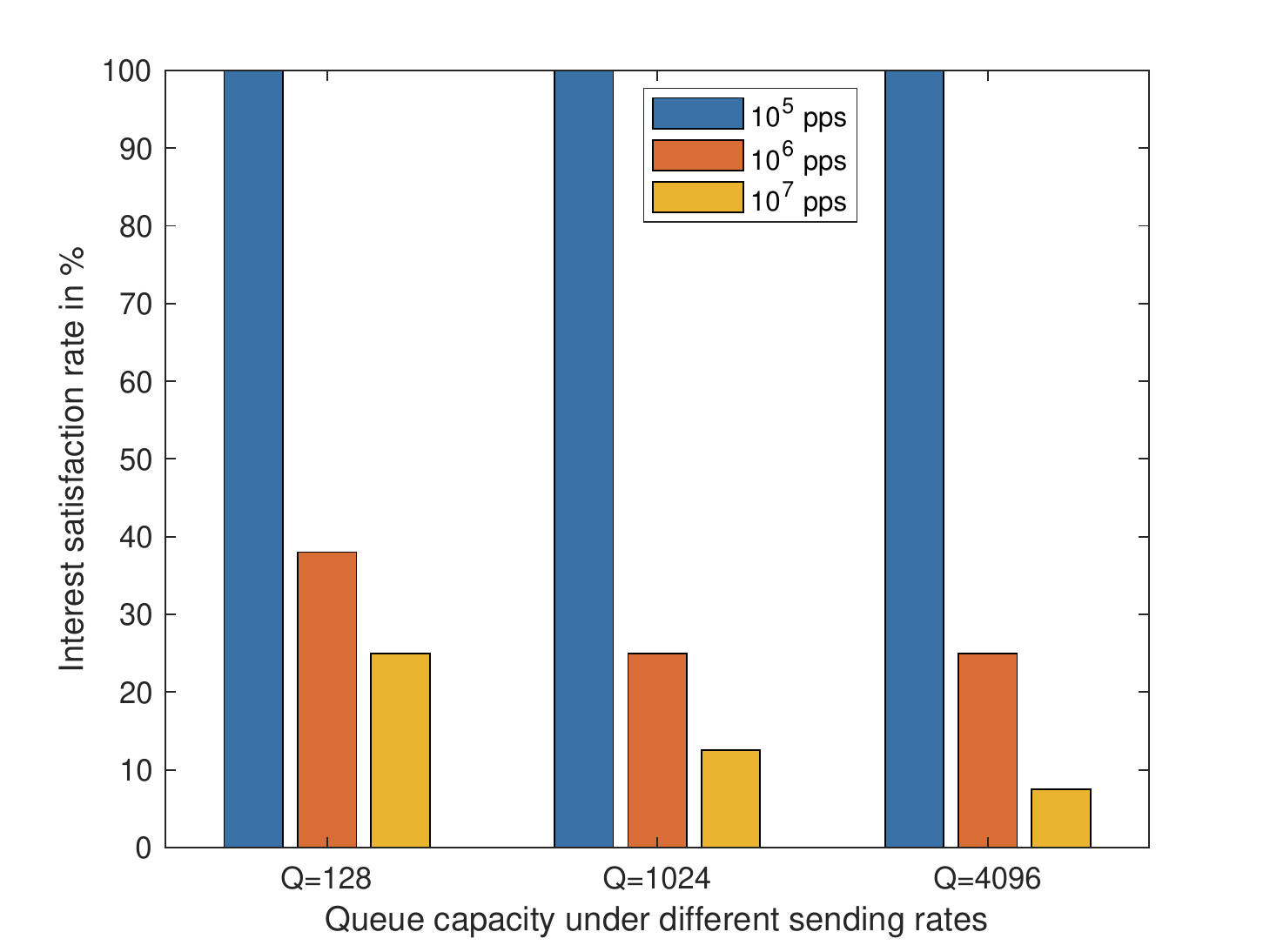}
		\label{fig:queuesize}
 	} 
	\quad
	\subfloat[Many Forwarding threads with a sending rate set to $10^6$ pps.]{
		    \includegraphics[width=.46\textwidth]{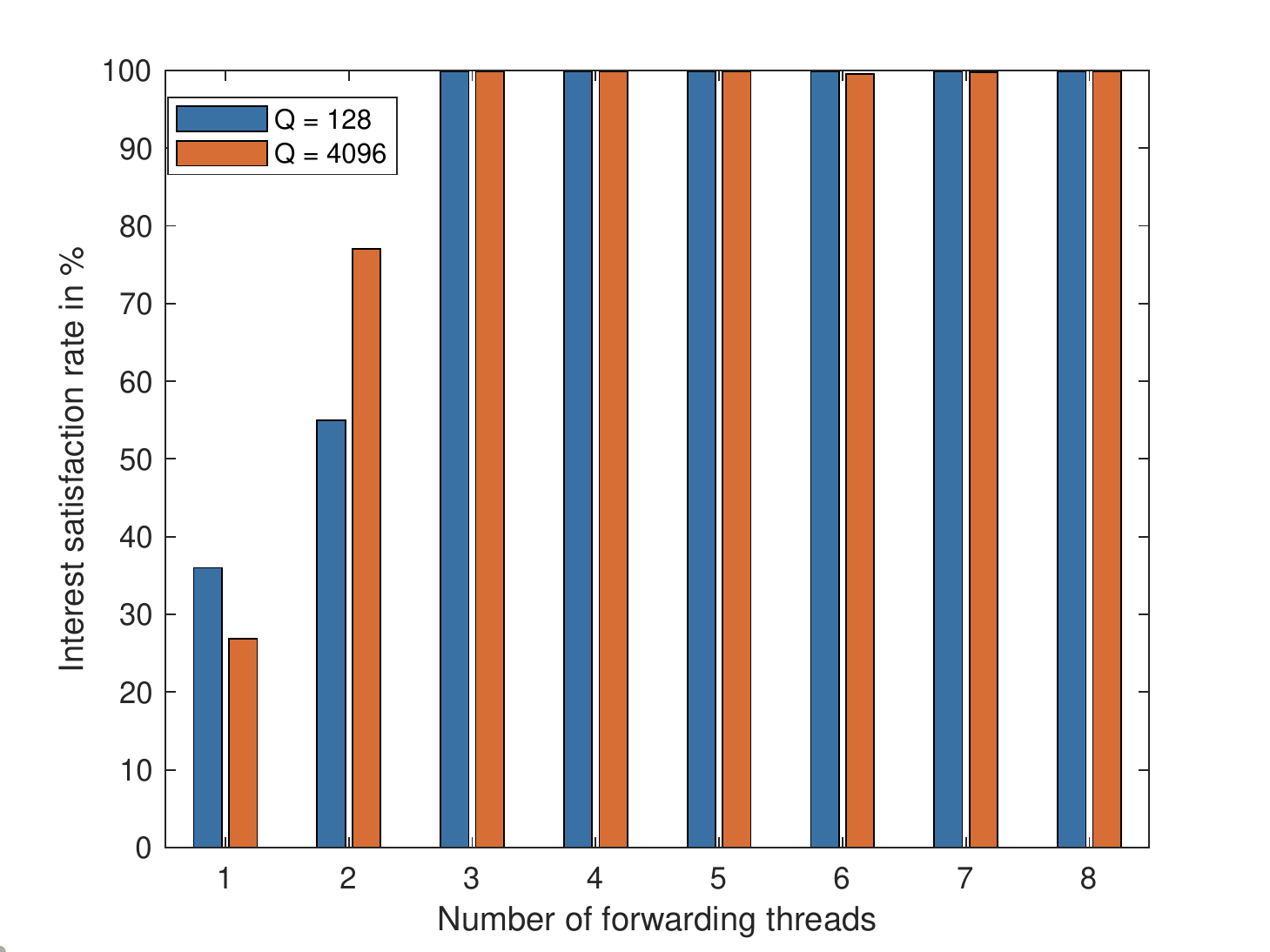}
		\label{fig:queueall}
	} 
	\caption{Exploration results of the Forwarding threads queues sizes.}
	\vspace{-0.6cm}
	\label{fig:queues}
\end{figure}

Fig.~\ref{fig:queuesize} indicates that at $10^5$ pps (blue), the Interest satisfaction rate is 100 \%. This means that the forwarder (with one forwarding thread) is capable of handling all packets at this sending rate ($10^5$ pps of packet size $1500$ bytes is equivalent to $1.2$ Gbps), under any queue size. However, under a faster sender rate (where a single forwarder shows signs of packet loss) we unexpectedly observed a better Interest satisfaction rate with a smaller queue (Q=$128$).
After a thorough investigation of the real implementation, we found out that the queues don't have proper management in terms of insertion and eviction policies that would give priority to Data over Interest packets. In the absence of such policy, more Interests would be queued while Data packets would be dropped resulting in Interests not being satisfied, thus lower performance (Interest satisfaction rate).
\emph{It is thus advised for the final implementation of the NDN-DPDK forwarder,  to use a queue capacity smaller than $128$ packets when the forwarder has a single forwarding thread and packets are sent at a fast rate.}

Similarly, we explore whether this observation remains true with more forwarding threads. In order to do that, we run SMC again on eight different models each with a different number of forwarding threads ($1$ to $8$) under a sending rate of $10^6$ pps ($1$ Interest per $1$ us) where a loss rate was observed in Fig.~\ref{fig:queuesize}. Then, we experimented with two queue capacities, namely $128$ and $4096$ packets. The results are reported in Fig.~\ref{fig:queueall}. The x Axis represents the number of forwarding threads while the y axis depicts the Interest satisfaction rate. 

 We observe that the queue size matters mainly in the case of a model with one and two forwarding threads. In fact, for a two threads model, a bigger queue size is preferred to maximize the performance, unlike when a single thread is used. As for the other six models, both sizes achieve almost 100 \% Interest satisfaction.
 This is due to the fact that three forwarding threads or more are capable of splitting the workload at $10^6$ pps and can pull enough packets from each queue with a minimum loss rate of 0.02 \% . \emph{This result stresses that, to avoid being concerned about a proper queue size, more threads are needed for handling a faster sending rate with minimum Interest loss.}

\subsubsection{NUMA placement, number of forwarding threads and packet name length.}
Another aspect to explore, is the impact of mapping the forwarding threads and/or NDN Faces to the two NUMA nodes (0, 1) under different sending rates and for multiple name lengths where Face 0 exchanges packets with the client and Face 1 with the server. To do that, we consider the four NUMA arrangements (P1), (P2), (P3) and (P4) in section \ref{sec:design} as well as the factors in Table~\ref{table:stochastic-factor-range} in the SMC analysis. 

In Figs.~\ref{s3} to \ref{f13}, each row represents experiments with similar packet name lengths \{small=3, medium=7, large=13\} and a queue capacity of 4096. The right-hand column indicates results for a faster sending rate of $2*10^6$ pps (500 ns interval) while the left-hand one shows results for a slower sending rate of $1.42*10^6$ pps (700 ns interval). 
The six figures includes four curves where each corresponds to the four NUMA arrangement options: P1 to P4.
\begin{figure}[htbp]
  \begin{minipage}[b]{0.52\linewidth}
    \centering
    \includegraphics[width=\linewidth]{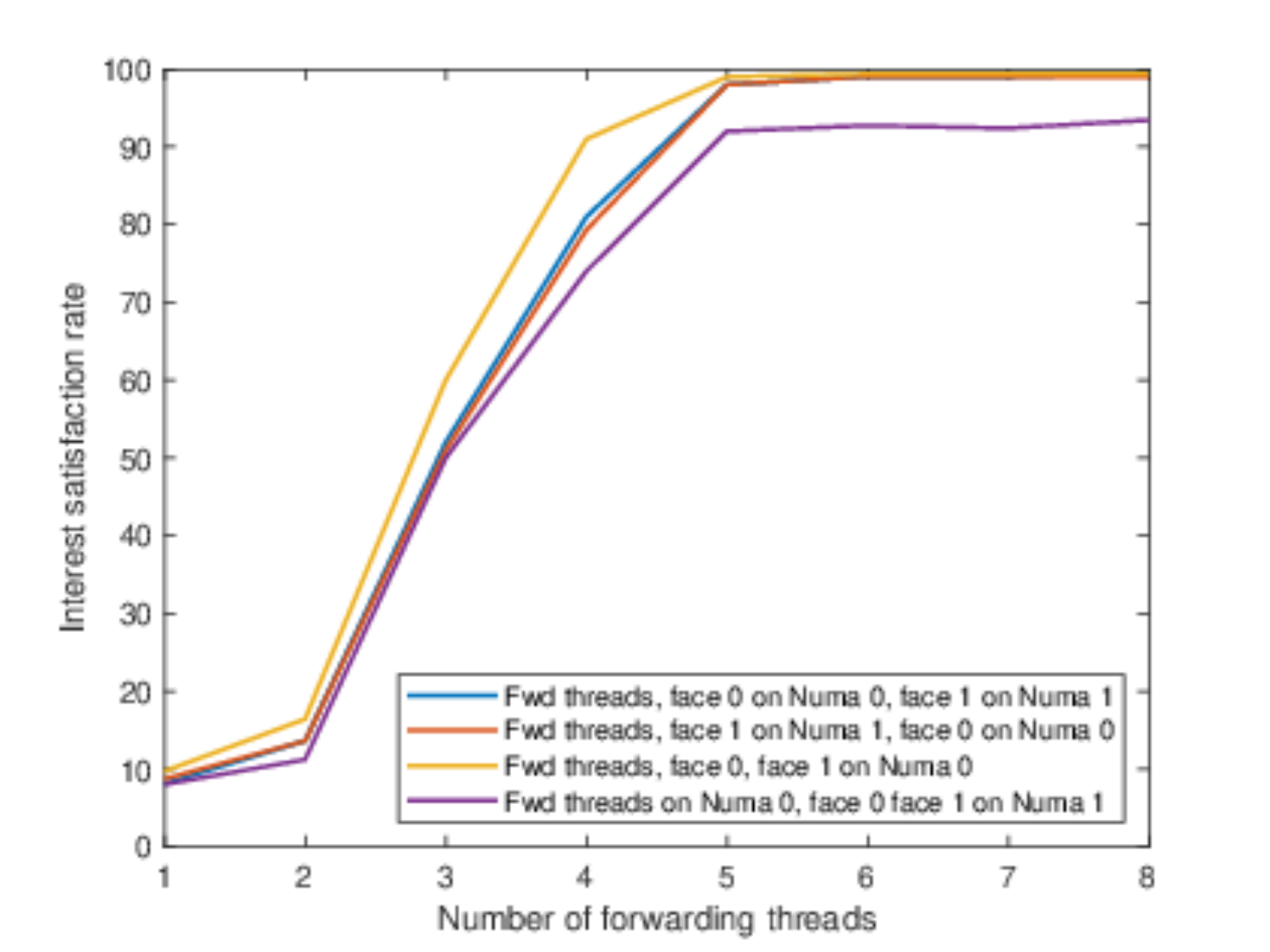}
    \caption{small names, 700 ns}\label{s3}
  \end{minipage}
  \begin{minipage}[b]{0.52\linewidth}
    \centering
    \includegraphics[width=\linewidth]{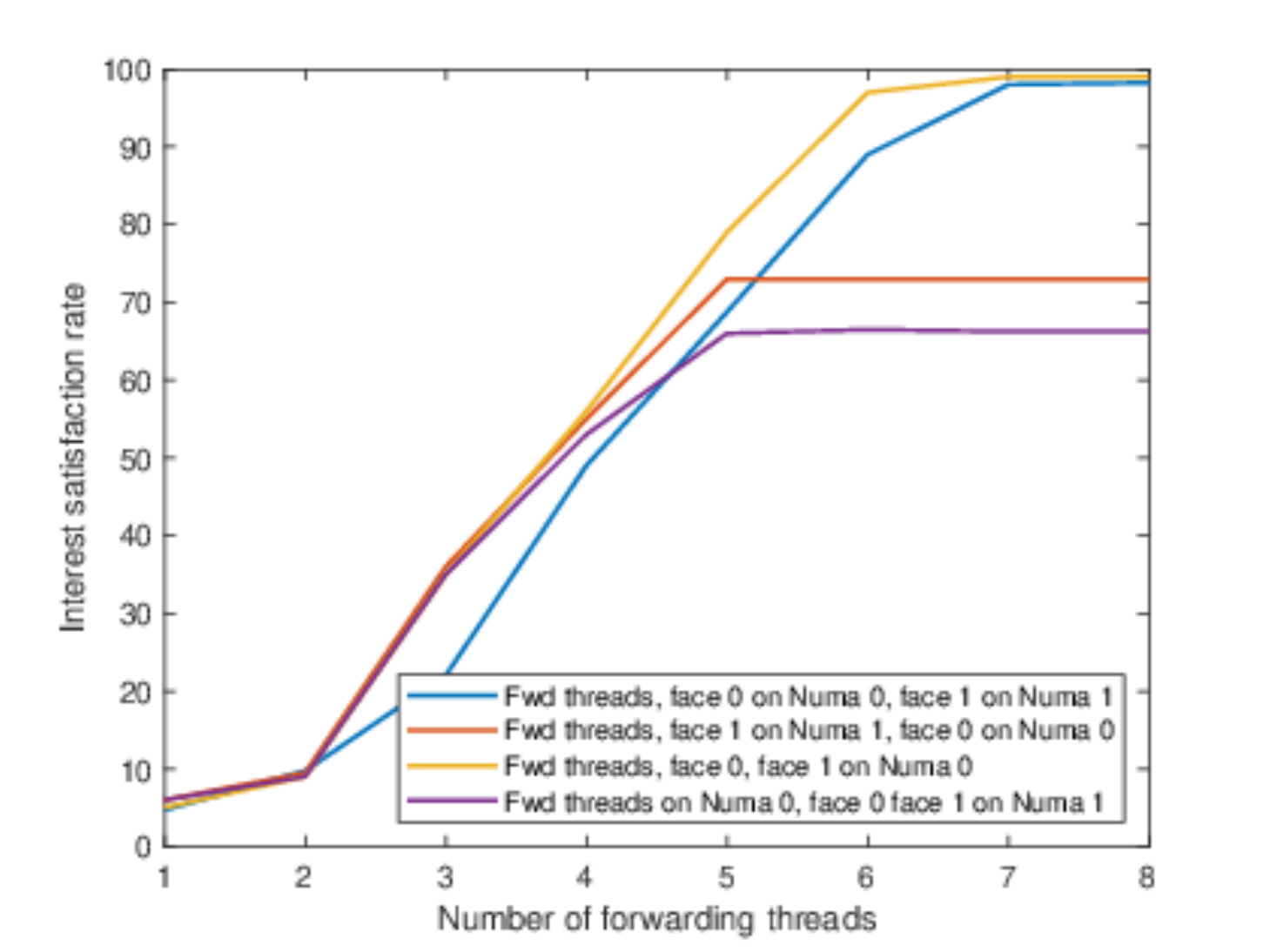}
    \caption{small names, 500 ns}\label{f3}
  \end{minipage}
    \begin{minipage}[b]{0.52\linewidth}
    \centering
    \includegraphics[width=\linewidth]{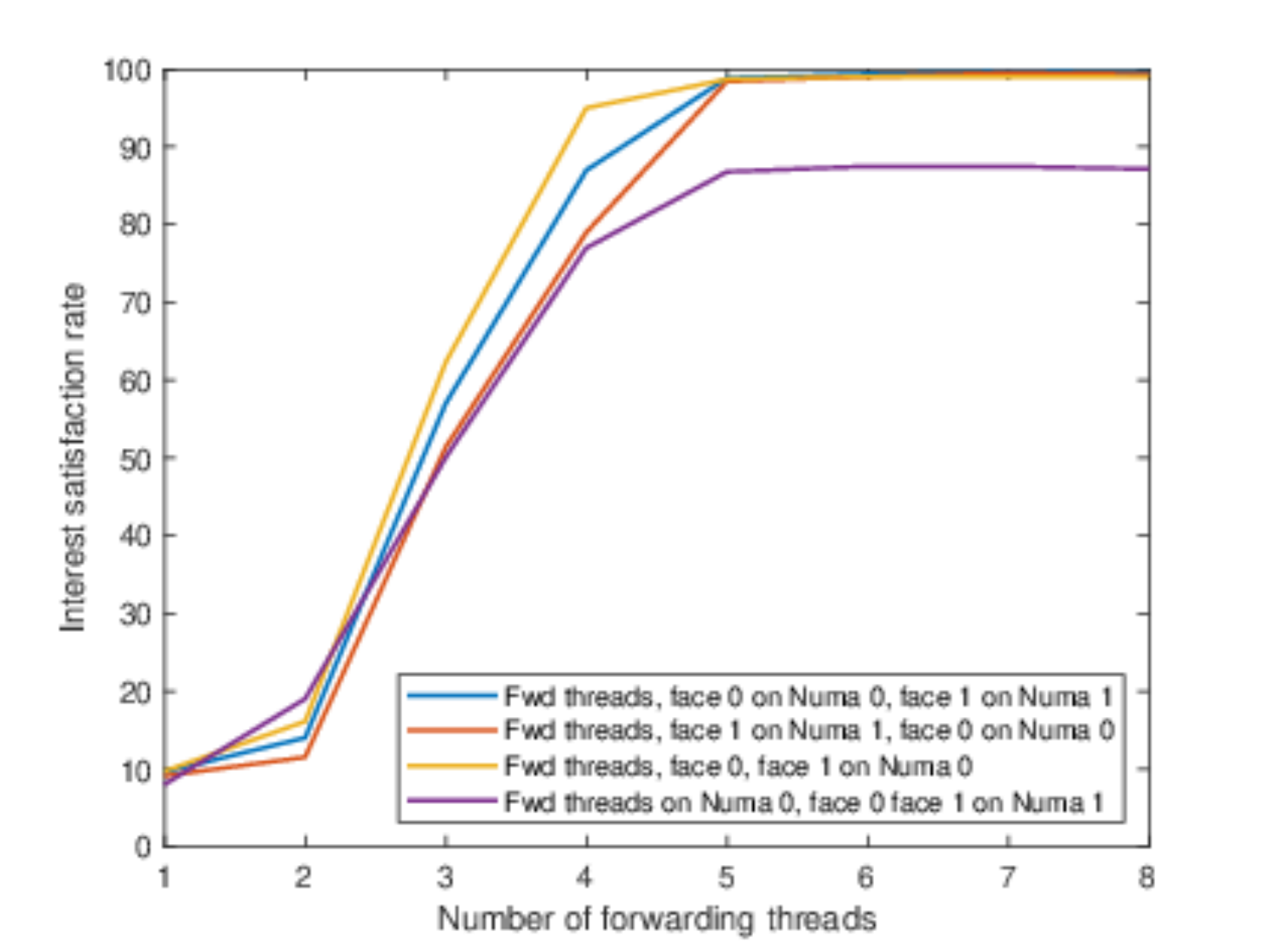}
    \caption{medium names, 700 ns}
    \label{s7}
  \end{minipage}
  \begin{minipage}[b]{0.52\linewidth}
    \centering
    \includegraphics[width=\linewidth]{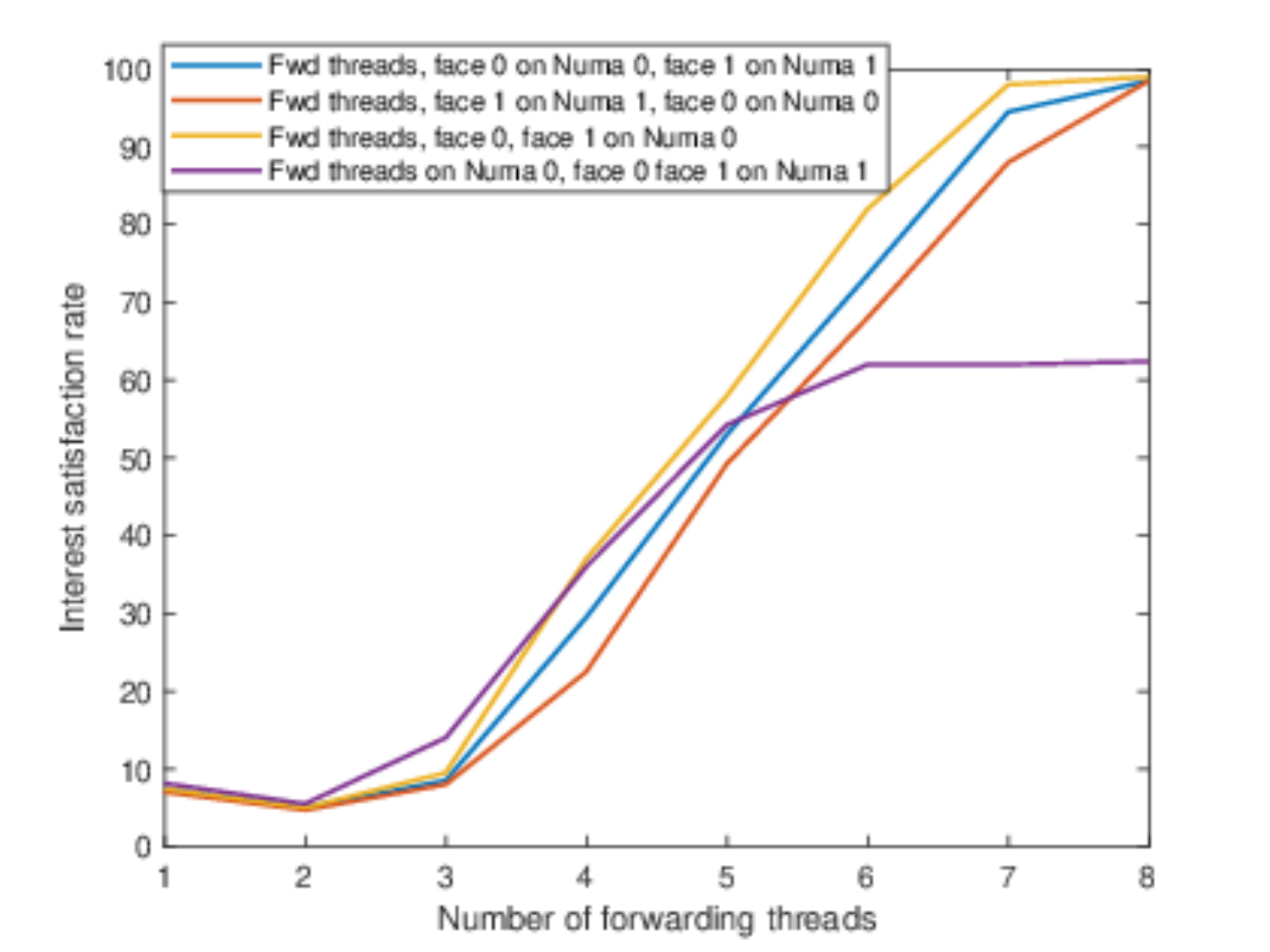}
    \caption{medium names, 500 ns}
    \label{f7}
  \end{minipage}
    \begin{minipage}[b]{0.52\linewidth}
    \centering
    \includegraphics[width=\linewidth]{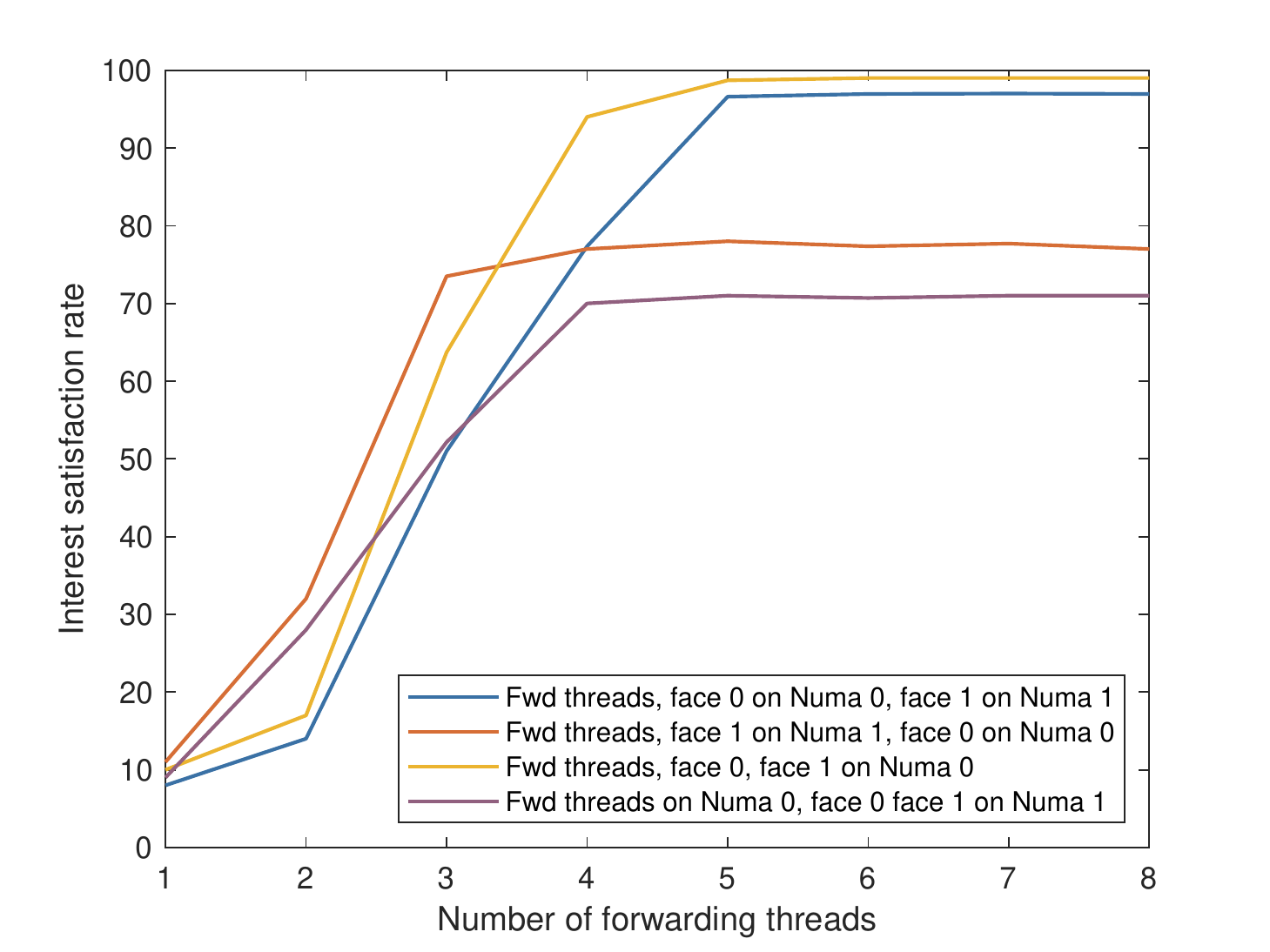}
    \caption{large names, 700 ns}
    \label{s13}
  \end{minipage}
  \begin{minipage}[b]{0.52\linewidth}
    \centering
	    \includegraphics[width=\linewidth]{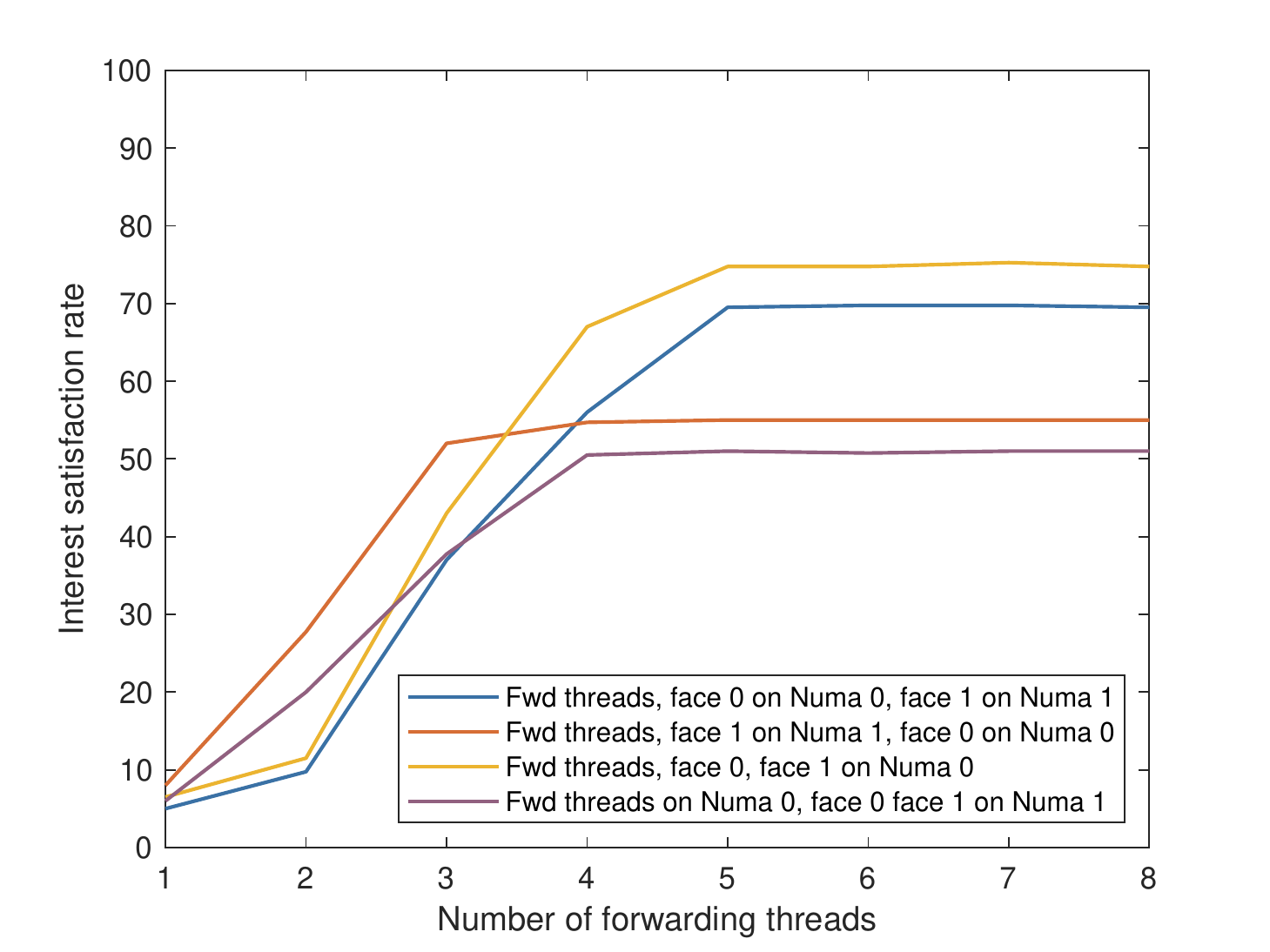}
    \caption{large names, 500 ns}
    \label{f13}
  \end{minipage}
\end{figure}

The six Figs.~\ref{s3} to \ref{f13} show that Interest satisfaction rates scale up with the increase of forwarding threads then reach a saturation plateau where adding more threads can no longer improve the performances.
Furthermore, with fewer forwarding threads, the loss rate is unavoidable and exceeds 80 \%. This is because the sending rate is faster than the forwarding threads processing capabilities causing their FIFO queues to saturate and start dropping packets frequently.
However, under a slower sending rate and packets with small, medium and large name lengths (3, 7, 13), Figs.~\ref{s3},~\ref{s7} and~\ref{s13} show that a maximum satisfaction rate of over 90 \% is achievable with only five forwarding threads.
Whereas when the client is generating packets faster at 2 Mpps, a saturation plateau of over $90$ \% is reached at six threads or more for small and medium names (Figs.~\ref{f3} and~\ref{f7}) and a plateau of slightly over $70$ \%, with five threads, for larger names (Fig.~\ref{f13}).
Also, Figs.~\ref{s3} and~\ref{s7} demonstrate that placing all processes (threads and faces) on a single NUMA (placement P1) outperforms the other three options. This observation is explained by the absence of inter-socket communication thus less timing overhead added such as in the case of the purple plot where packets are crossing NUMA boundaries twice from Face 0 to the forwarding threads then through Face 1 and back (placement P4).

\vspace{-0.03cm}
Figs.~\ref{f3} and~\ref{f7} show the impact of increasing the sending rate on packets with smaller names. In this case, it is preferred to also position all the processes on one NUMA such as the case of the yellow plot of the P1 series because NUMA boundary crossing usually downgrades the performance. In fact, the difference between no NUMA crossing and the double crossing (yellow and purple series respectively) is approximately 30 \% loss rate with more than five threads. The second best option P2 which is placing the forwarding threads on the NUMA receiving Interest packets with Face 0 (NUMA hosting the Ethernet adapter that receives Interests from the Client). However, when the number of threads is not in the saturation zone and the threads get overworked and start to loose packets, it is recommended to opt for placement P3. \emph{Based on these results, we recommend that for small to medium names, to use a maximum of eight threads but no less than five arranged as in placement P1 for optimum performances under a slower or a faster sending rate}. 

With a larger name however, Fig.~\ref{s13} depicts an unexpected behaviour when using three threads or less. In this case, placing the forwarding threads on the same NUMA as Face 1 (which is the Ethernet adapter connected to the server and receives Data packets), surpasses the other three options. Our explanation is that since forwarding threads take longer times to process incoming packets due to their longer name and timely lookup, particularly for Interests as they are searched by names inside the two tables (PCCT and FIB) rather than a token such as the case for Data packets. Placing the forwarding threads with the Data receiving Ethernet adapter connected to Face 1, has the potential to yield better results by quickly processing packets after a quick token search especially when the workload is bigger than the threads' processing capacity. When the sending rate is increased, the same results are observed in Fig.~\ref{f13} for a similar name length but with a decrease in performance. \emph{Thus, we recommend for larger names to use NUMA arrangement P3 only when the number of forwarding threads is less than three regardless of the sending rate (not advised due to high loss rate).}

\paragraph{}

\bibliographystyle{splncs04}
\bibliography{main}
\end{document}